\documentclass[%
reprint,
superscriptaddress,
amsmath,amssymb,
aps,
prapplied,
floatfix,
]{revtex4-2}

\usepackage{graphicx}
\usepackage{dcolumn}
\usepackage{bm}
\usepackage{enumitem}
\usepackage{array}

\usepackage{tabularx}
\usepackage{array}
\usepackage{float} 
\usepackage{indentfirst} 
\usepackage{hyperref}
\hypersetup{colorlinks,linkcolor=blue,citecolor=blue, urlcolor=blue}
\usepackage{color}
\usepackage{amsmath} 
\usepackage{physics}
\usepackage{mathtools}
\usepackage{mleftright}
\usepackage{chemformula}
\usepackage{soul}

\newif\ifshowchanges
\showchangesfalse   

\ifshowchanges
    \sethlcolor{yellow}
\else
    \renewcommand{\hl}[1]{#1}
\fi

\usepackage{enumitem}

\usepackage[normalem]{ulem}
\usepackage{array}
\usepackage{minted} 
\usepackage{indentfirst}
\usepackage{multirow}
\usepackage[most]{tcolorbox}
\newtcolorbox{designprinciplebox}{
  colback=gray!5,
  colframe=black,
  boxrule=0.5pt,
  arc=2pt,
  left=6pt,
  right=6pt,
  top=6pt,
  bottom=6pt
}
\usepackage{xcolor}

\begin{document}
\preprint{APS/123-QED}

\title{From projected density of states to spatially resolved electronic descriptors in III-nitride–nanocluster photocatalysts: A database-driven \textit{ab initio} study}
\title{Region-Resolved Interface Descriptors in III-nitride-nanocluster photocatalysts}
\author{Shuaishuai Yuan}
\affiliation{Department of Physics, McGill University, Montr\'eal, Qu\'ebec, Canada}
\author{Gunther G. Andersson}
\affiliation{College of Science and Engineering, Flinders University, Adelaide, South Australia, Australia}
\author{Gregory F. Metha}
\affiliation{Department of Chemistry, University of Adelaide, Adelaide, South Australia, Australia}
\author{Zetian Mi}
\affiliation{Department of Electrical Engineering and Computer Science, University of Michigan, Ann Arbor, Michigan, USA}
\author{Hong Guo}
\affiliation{Department of Physics, McGill University, Montr\'eal, Qu\'ebec, Canada}
\date{\today}





\begin{abstract}
Understanding how nanocluster cocatalysts modify the electronic structure of III-nitride surfaces is central to the rational design of efficient photocatalytic interfaces.
Here, we \hl{introduce a region-resolved first-principles framework} for analyzing GaN-, InGaN-, and ScGaN-based slabs decorated with six-atom elemental nanoclusters.
Using a region-resolved projected local density of states (PLDOS) analysis, we \hl{separate the electronic response of nanocluster-covered regions from that of uncovered semiconductor surface regions}, revealing that semiconductor--nanocluster interfaces behave as laterally heterogeneous electronic systems. 
In this picture, nanocluster-covered regions are associated with local band bending, charge injection, and nanocluster-induced gap states, whereas uncovered nitride regions retain surface-localized states and in-plane dipoles \hl{that may influence} adsorbate interactions and water-activation pathways. \hl{These distinct regional responses motivate the definition of region-resolved interface descriptors, which are further integrated into an Interface Score: a physics-informed comparative metric that combines injection, activation, and interfacial-coupling contributions to rank heterogeneous model interfaces on a common descriptor basis.} Machine-learning models trained on global and region-resolved descriptors are used primarily as interpretability tools to assess their correlation with single-H adsorption energetics. The results show that spatially averaged descriptors capture broad adsorption-energy trends, while region-resolved descriptors expose complementary local information related to interfacial electrostatics and charge separation. Together, this work establishes \hl{a generalizable descriptor framework for moving beyond spatially averaged electronic-structure analysis and for bridging first-principles calculations with experimental interpretation of heterogeneous photocatalytic interfaces.}
\end{abstract}

\maketitle
\section{Introduction}
Photocatalytic hydrogen production offers a promising route toward sustainable energy conversion by directly harnessing solar energy to drive water splitting reactions \cite{davis2018, guan2023, hisatomi2019, johnson2025, pinaud2013, rodriguez2014}. Among candidate materials, III-nitride semiconductors such as GaN and its InGaN alloys are particularly attractive due to their chemical stability, tunable band gaps, and favorable band-edge alignment with water redox potentials \cite{wang2019, xiao2022, zhou2023, zhou2024}. Recent experimental advances, including record solar-to-hydrogen efficiencies achieved in InGaN nanowire systems \cite{zhou2023}, highlight the potential of III-nitrides as a robust platform for photocatalytic hydrogen evolution. Despite this progress, the microscopic electronic mechanisms \cite{kamat2012, li2019, vandeelen2019, xu2023, zhang2012} by which cocatalysts interact with III-nitride surfaces, and the manner in which these interactions govern charge separation and surface reactivity, remain incompletely understood.

Nanocluster cocatalysts are increasingly employed to enhance photocatalytic performance by providing active sites for surface redox reactions and facilitating charge transfer \cite{doan2023, klumpers2024, krishnan2017, liu2024, manna2023}. At the nanoscale, however, cocatalysts do not merely act as passive reaction centers. Their interaction with the semiconductor surface can induce local band bending, nanocluster-induced gap states, and charge redistribution, all of which depend sensitively on cocatalyst chemistry, size, and interfacial bonding \cite{zhang2012, li2017, xu2023}. As a result, nanocluster-decorated III-nitride surfaces should be viewed not as homogeneous catalytic planes, but as laterally heterogeneous electronic interfaces in which spatially distinct regions play different functional roles.

From a theoretical perspective, establishing transferable design principles for semiconductor–cocatalyst interfaces remains challenging. Existing first-principles studies are often limited to isolated case studies, with substantial variability in surface orientation, termination, slab construction, cocatalyst configuration, and analysis protocols \cite{cohen2008, mou2023, gusarov2024, lin2024, samanta2022}. This diversity hinders systematic comparison across systems and obscures general trends linking interfacial electronic structure to photocatalytic performance. A second, equally fundamental limitation lies in how interfacial electronic structure is analyzed. Commonly used tools such as projected density of states and charge-density plots provide spatially averaged information, obscuring local variations in interfacial electronic structure. This limitation is particularly severe in III-nitride-nanocluster systems, where strong spontaneous polarization, dense surface states, and weak electrostatic screening give rise to distinct interfacial electrostatic landscapes between nanocluster-covered and uncovered regions. Spatially averaged descriptors are therefore insufficient to capture the physics of charge separation and interfacial transport \cite{zhang2012}.

To move beyond case-specific interpretations while capturing interfacial physics, a framework is needed that simultaneously resolves spatially heterogeneous electronic structure and enables systematic comparison across chemical space. Data-driven approaches, when grounded in physically meaningful, spatially resolved descriptors, offer a powerful route to quantify how electronic structure, electrostatics, and chemical bonding collectively influence interfacial reactivity \cite{shi2021, mai2022, toyao2020, mou2023, xu2024}. 
In this work, we seek to \hl{develop a computational framework for extracting electronic-structure design insights} for nanocluster cocatalysts on III-nitride surfaces.
We construct a systematic first-principles dataset of GaN-based nanocluster interfaces spanning a wide range of cocatalyst elements and alloy compositions. To investigate the resulting interfacial electronic structure, we introduce a region-resolved projected local density of states (PLDOS) framework that extracts band bending, surface states, and charge redistribution across different spatial regions. Physically motivated descriptors derived from this analysis are then used to assess their impact on hydrogen adsorption energetics, machine-learning models serve as a quantitative tool to evaluate descriptor importance and generalize trends across chemical space. 
Together, this approach \hl{provides a framework for incorporating charge injection, surface activation, and their coupling into ab initio and data-driven design} of semiconductor–nanocluster interfaces.
\hl{Importantly, the region-resolved framework developed here is intended as a complementary, screening-level approach rather than a substitute for explicit kinetic reaction-barrier calculations. For heterogeneous semiconductor--nanocluster interfaces, the active site and reaction pathway are often not uniquely defined and may vary among nanocluster-covered regions, uncovered substrate regions, and interfacial boundary regions. Rather than focusing on a single elementary reaction step, the present approach maps the interfacial electronic structure onto physically interpretable ground-state descriptors, thereby enabling systematic comparison across diverse photocatalytic interfaces.}

\hl{Three key design insights emerge from the material systems studied here}: (i) Lateral heterogeneity is \hl{important}, as spatial separation between covered and uncovered regions drives complementary photocatalytic functions; (ii) Interfacial electrostatics govern transport, with band bending and dipoles playing roles that extend beyond what is captured by conventional adsorption scaling relations; and (iii) Substrate selection defines the operational baseline, establishing the fundamental electrostatic landscape within which nanocluster chemistry fine-tunes the local electronic structure.

The rest of this paper is organized as follows. Sec.~\ref{sec:ap_methods} describes the computational methods, including first-principles calculations and machine-learning approaches used to construct the dataset. Sec.~\ref{sec:Results} presents the results, including an overview of the dataset and model systems, the development of region-resolved descriptors from PLDOS analysis, the construction of the Interface Score, and data-driven feature attribution for physical interpretability. Sec.~\ref{sec:Discussion} discusses the resulting design \hl{insights} and data-driven \hl{interpretations}. Sec.~\ref{sec:conclusion} concludes the paper.

\section{Methods}
\label{sec:ap_methods}

In this section, we describe the computational framework used to construct the dataset and extract region-resolved electronic descriptors, including first-principles calculations and machine-learning methods.

\subsection{DFT calculations}
\label{sec:ap_dft}

 All DFT calculations in this study were performed within the Vienna \textit{Ab initio} Simulation Package (VASP) \cite{kresse1993, kresse1994, kresse1996, kresse1996a}, utilizing the generalized gradient approximation (GGA) functional by Perdew, Burke, and Ernzerhof (PBE) \cite{Perdew1996}. Projector-augmented-wave (PAW) potentials \cite{Blochl1994, Kresse1999a}  were selected according to VASP’s default recommendations from the PBE PAW dataset (version 54). A plane-wave energy cutoff of 550 eV was applied in all calculations. Additionally, all electronic structure calculations were spin polarized and Gaussian smearing of 0.05 eV was used.

The initial configurations of GaN with space group P6$_3$mc were obtained from the materials project \cite{Jain2013}, and relaxations are performed using a \hl{$\Gamma$-centered 8 $\times$ 8 $\times$ 5} k-point mesh. For all supercell calculations, we used only the $\Gamma$-point for Brillouin-zone sampling. The PBE-D3 functional \cite{Grimme2010} was employed for the exchange–correlation interactions, as it provides improved accuracy in reproducing experimental lattice parameters. 

To model InGaN and ScGaN alloys, we constructed a 4$\times$4$\times$4  supercell in which 25\% of the Ga sites were substituted by In or Sc atoms.
The optimized special quasi-random structure (SQS) was generated using the sqsgenerator package \cite{gehringer2023, vandewalle2013, wei1990, zunger1990}  to more accurately simulate random atomic distributions under experimental conditions. In this study, we investigated the (110) facets of GaN, In$_x$Ga$_{1-x}$N ($x$=0.25), and Sc$_x$Ga$_{1-x}$N ($x$=0.25). The surface slabs were generated using the pymatgen package \cite{Ong2013}. 
\hl{
The alloy composition of 25\% was selected as a representative intermediate concentration, consistent with experimentally reported In contents of 9\%-40\% in highly efficient GaN-based nanowire photocatalysts} \cite{zhou2023}. \hl{Since GaN-based nanowires typically grow along the c axis, their dominant catalytic interfaces are the nonpolar sidewall facets, such as (100) and (110), which expose both metal and nitrogen atoms. We selected the (110) surface as a representative model system. The computational framework developed here is readily applicable to other crystallographic orientations and related material systems.}

Nanoclusters were selected from the Quantum Cluster Database (QCD), which contains low-energy cluster structures for 55 elements, with cluster sizes ranging from 3 to 55 atoms, computed using density functional theory (DFT) \cite{manna2023}. From this database, we extracted the most stable 6-atom nanoclusters for each element of interest. \hl{The six-atom cluster size and the lowest-energy QCD structures were used as a standardized baseline that enables systematic comparison of cocatalyst chemistry across the periodic table, rather than as an exhaustive sampling of all experimentally possible cluster sizes, orientations, or morphologies.}

The semiconductor–nanocluster interfaces were modeled using a slab geometry (Fig.~\ref{fig:Dataset}). Each slab contains 518 atoms and includes a vacuum region exceeding 20 {\AA} to prevent interactions between periodic images along the out-of-plane direction. The bottom two atomic layers were fixed to mimic bulk termination, while the remaining atoms were allowed to relax. Six-atom nanoclusters composed of different elements were subsequently deposited on the exposed surface to systematically probe how cocatalyst chemistry modifies the interfacial electronic structure. Because fully relaxing these large semiconductor–nanocluster systems at the DFT level is computationally prohibitive, we employed the pretrained MACE foundation model (mace-mpa-0-medium) \cite{batatia2022} to perform the structural relaxations, \hl{with the optimization run until the maximum atomic force was below 0.01 eV/\AA}. The bottom two layers of the slab were fixed, while all other atoms were allowed to relax along the vertical direction. The in-plane lattice parameters were constrained to their bulk values. 

\hl{Due to the vast combinatorial space of possible adsorption sites across the laterally heterogeneous surface, H adsorption was restricted to a representative site on the nanocluster (Region A), which serves as the primary electron-injection and hydrogen-reduction center. Specifically, the H atom was initially positioned above the nanocluster atom with the highest z-coordinate in the previously relaxed structure, after which  the H atom was relaxed using the pretrained MACE foundation model. While a more comprehensive exploration of adsorption sites and reaction intermediates (e.g., H$_2$O or OH$^-$
 adsorption in Region B) would be desirable, it lies beyond the scope of the present work and represents a natural extension of the proposed framework.}

We did not perform system-specific fine-tuning of the MACE model, as the nanoclusters span the entire periodic table and each relaxed final geometry is subsequently evaluated with full DFT. The MACE-relaxed structures were then used for static electronic calculations employing the  DFT parameters described above. \hl{For Bader charge analysis, the all-electron-like charge density was reconstructed by enabling \texttt{LAECHG = .TRUE.}, and the resulting charge density was used to partition atomic charges.} Dipole corrections were applied along the $z$-direction to eliminate spurious interactions between periodic images. 
\hl{We note that while the pretrained MACE foundation model enables the high-throughput evaluation of these massive 518-atom supercells, the resulting local nanocluster geometries can exhibit deviations from the \textit{ab initio} local minima. Full structural relaxation at the DFT level typically requires hundreds of ionic steps per system, which is beyond our computational resources. Although we believe the framework and overall insights achieved in the current work are not limited to the precise microscopic atomic arrangement, this work primarily serves to demonstrate an innovative \textit{ab initio} framework, helping to provide design ideas and screen new systems as a first pass. For the exact catalytic evaluation of specific high-performing candidates, full DFT structural optimization remains necessary.}

\subsection{Machine learning methods}
\label{sec:ap_ml_method}

All machine-learning workflows used Python (3.11.8) with scikit-learn (1.6.1). The dataset comprised 127 adsorption cases, and two descriptor sets were constructed: (i) 30 global descriptors (hydrogen PDOS band properties—centers/widths/edges and integrated weights near VBM/gap/CBM—host band‑edge positions, elemental Bader charges, work function/vacuum alignment, and group/period identifiers), etc. (Details see Table~\ref{tab:global}); and (ii) 23 region-resolved/interface descriptors from spatial PLDOS (NIGS, Bader Charge Difference between surface/subsurface layers, surface-state peak energies/widths, and layer-resolved band-bending metrics), etc. (Details see Table~\ref{tab:iface}). We used an 80/20 holdout split (101 train/validation, 26 test; random\_state=2). \hl{
Additional robustness tests, including 50 repeated random 80/20 splits, leave-one-element-out validation, leave-one-substrate-out validation, and random-label controls, are reported in Appendix}~\ref{app:ml_validation_controls}. Hyperparameters for six models (RF, ET, GBR, HGB, SVR, LASSO) were optimized with RandomizedSearchCV (40 draws) using repeated K‑fold CV (8 folds × 3 repeats; random\_state=42) on the training portion only. Scaling (StandardScaler) was used for models that require it (SVR/LASSO), while tree models used raw features. Performance was reported via $R^2$ and RMSE with parity plots. PCA (after standardization) was used for latent‑space visualization. SHAP values (TreeExplainer) were computed for the best models per descriptor set (GBR for global; ExtraTrees for interface) to assess descriptor importance in predicting H adsorption energies. Core libraries included NumPy, SciPy, Matplotlib, and scikit‑learn to ensure reproducibility. Hyperparameter ranges and best settings are listed in Appendix~\ref{app:hyperparams}.

\section{Results}
\label{sec:Results}

\subsection{Overview of the dataset and model systems}

\begin{figure*}
\centering
\includegraphics[scale=1]{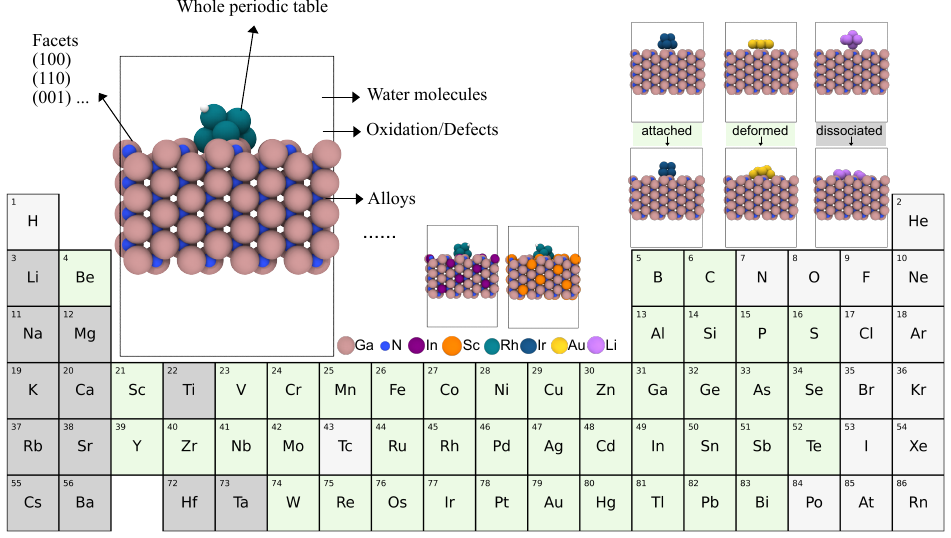}
\caption{Schematic of the constructed \textit{ab initio} database comprising GaN, InGaN, and ScGaN slabs decorated with six-atom elemental nanoclusters spanning the periodic table. The framework can accommodate additional variations such as facets, alloying, oxidation or defect states, and water adsorption. Upon relaxation, the nanoclusters exhibit three structural motifs: attached (marked in green), deformed (marked in yellow), and dissociated (marked in dark gray). Elements shown in light gray do not form stable six-atom nanoclusters on GaN-based surfaces.}
\label{fig:Dataset}
\end{figure*}

Motivated by the record solar-to-hydrogen efficiencies achieved in InGaN nanowire photocatalysts \cite{zhou2023}, we perform an \textit{ab initio} characterization of the interfacial electronic structure of nanocluster-decorated III-nitrides. InGaN is a prototypical semiconductor with a tunable band gap, strong light absorption, and excellent chemical stability, making it a promising platform for photocatalytic water splitting \cite{fathabadi2025, wang2019, zhou2023, zhou2024}. A central question addressed here is how alloy composition and nanocluster loading jointly shape interfacial band alignment, electrostatics, and charge transfer, which are key ingredients for \hl{providing useful trends and insights beyond empirical optimization.}

We focus on the ground-state electronic structure of group III nitride surfaces modified with nanoclusters, which is a fundamental component for controlling interfacial charge separation and provides a clear benchmark for more complex photocatalytic modeling \cite{samanta2022, mou2023}. Although the schematic in  Fig.~\ref{fig:Dataset} illustrates a broader range of possible facets and environments, the present dataset is limited to the (110) orientation, which is nonpolar along the surface normal but polar in-plane \cite{zhong2020}. Future extensions will incorporate additional polar and semi-polar surfaces, as well as water-covered and defect-containing configurations \cite{alqahtani2016, krishnan2017, xiao2023, zhong2020, fathabadi2025}.

Fig.~\ref{fig:Dataset} summarizes the constructed dataset. All stable six-atom nanoclusters were deposited on GaN-based (110) slabs and subsequently relaxed using a consistent computational workflow (see Sec.~\ref{sec:ap_methods}). After relaxation, the nanoclusters adopt three distinct geometrical motifs: attached, deformed, and dissociated, \hl{which are color coded in the periodic table shown in} Fig.~\ref{fig:Dataset}. The attached and deformed configurations, which correspond to stable interfacial bonding without fragmentation, form the focus of the present spatially resolved electronic-structure analysis. \hl{Overall, the final dataset comprises 127 unique stable adsorption configurations with complete descriptor sets, obtained by relaxing 55 candidate six-atom nanoclusters on 518-atom GaN, In$_{0.25}$Ga$_{0.75}$N, and Sc$_{0.25}$Ga$_{0.75}$N (110) supercells and retaining attached or deformed motifs for subsequent DFT calculations and descriptor-based analysis.}

\subsection{From PDOS to region-resolved PLDOS}
\label{sec:PLDOS}

\begin{figure*}
\centering
\includegraphics[scale=0.8]{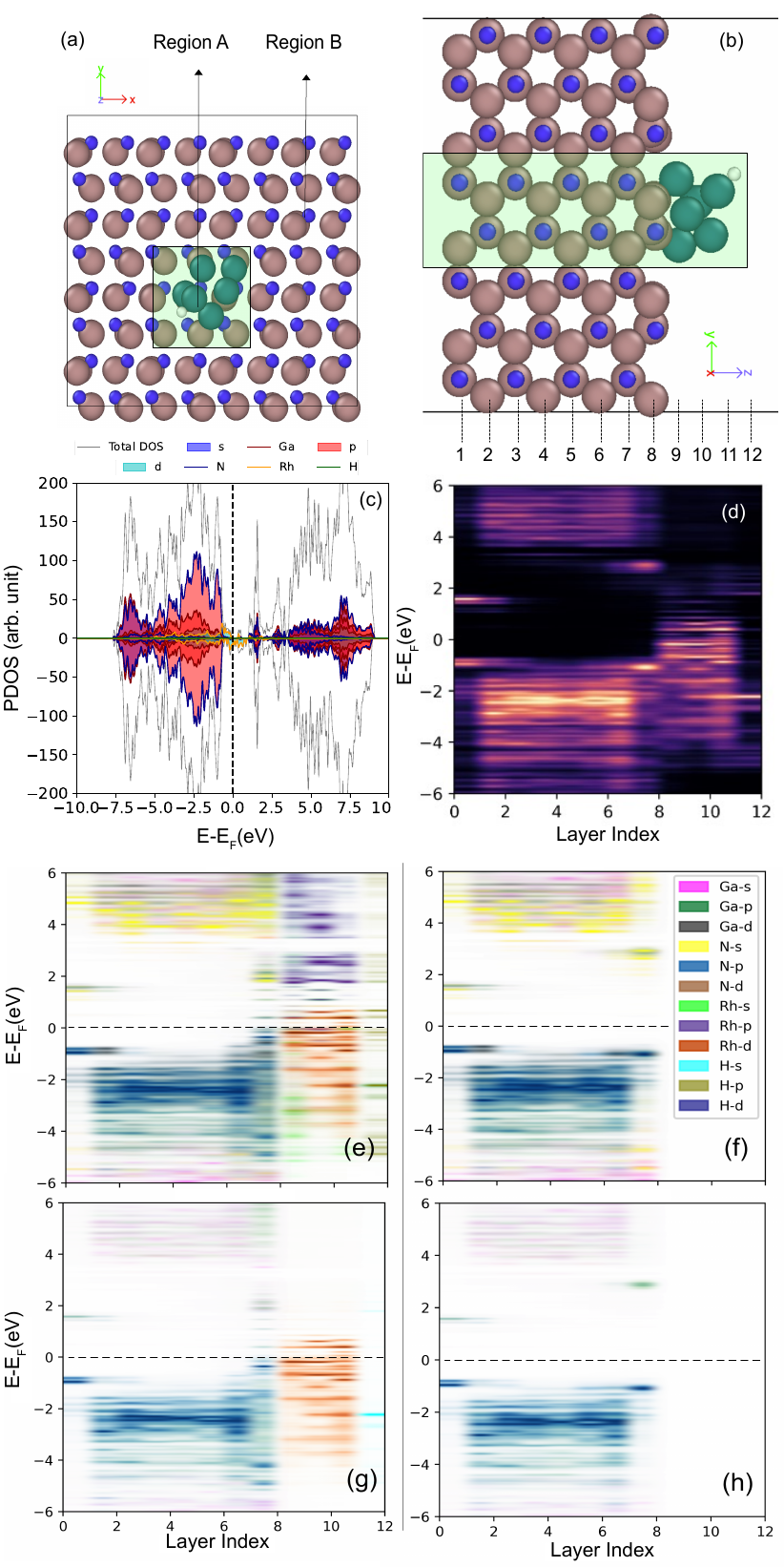}
\caption{From PDOS to region-resolved PLDOS for a six-atom Rh nanocluster on a GaN (110) slab with a single hydrogen atom adsorbed on the nanocluster.  (a,b) Top and side views of the relaxed structure, defining Region A (beneath the nanocluster) and Region B (uncovered surface). (c) PDOS of the Rh-decorated slab, illustrating overall density-of-states overlap but limited spatial resolution. (d) Layer-resolved projected local density of states (PLDOS) revealing band alignment and charge redistribution across the interface. (e,f) Region-resolved PLDOS for Regions A and B, normalized per orbital channel to emphasize spatial variations across layers. (g,h) PLDOS with element-wise normalization highlighting dominant orbital contributions in each region.}
\label{fig:DOS_to_PLDOS}
\end{figure*}

To expose the spatially heterogeneous electronic structure induced by nanocluster adsorption, we examine a representative case of a six-atom Rh nanocluster on a GaN (110) slab with a single hydrogen atom adsorbed on the nanocluster. Figures~\ref{fig:DOS_to_PLDOS}(a,b) show the top and side views of the relaxed configuration. The surface is partitioned into two regions: Region A, directly beneath the nanocluster, and Region B, the uncovered GaN surface. This partitioning enables direct comparison between nanocluster-covered and bare surface regions, allowing local electronic responses to be resolved.

Fig.~\ref{fig:DOS_to_PLDOS}(c) shows the PDOS of the Rh-decorated slab. While the slab PDOS captures overall Rh–GaN hybridization, its spatially averaged nature obscures differences between nanocluster-covered and uncovered regions, masking local band bending and surface-state redistribution essential for interfacial charge transport.

To recover spatial information, we evaluate the projected local density of states (PLDOS) resolved along the surface normal. The resulting layer-resolved map [Fig.~\ref{fig:DOS_to_PLDOS}(d)] clearly delineates the GaN substrate layers (layers 1–8), the Rh nanocluster (layers 9–11), and the adsorbed H atom (layer 12), revealing distinct interfacial band alignment and nanocluster-induced band bending. However, because the layer-resolved PLDOS in Fig.~\ref{fig:DOS_to_PLDOS}(d) is averaged over the entire lateral extent of each atomic layer, it still mixes contributions from nanocluster-covered and uncovered regions. As a result, the magnitude and spatial localization of band bending are not fully resolved. This distinction becomes much clearer in the region-resolved PLDOS analysis introduced below, where nanocluster-induced band bending is shown to be strongly localized in Region A beneath the nanocluster.

To go beyond both spatially averaged PDOS and laterally averaged layer-resolved PLDOS, we perform a region-resolved PLDOS analysis for Regions A and B.
Figs.~\ref{fig:DOS_to_PLDOS}(e,f) show PLDOS profiles normalized within each orbital channel (independent scaling), which highlights spatial variations and band bending even for low-intensity orbitals. In contrast, Figs.~\ref{fig:DOS_to_PLDOS}(g,h) use element-wise normalization (relative scaling) to emphasize the relative magnitude of dominant orbital contributions, revealing the primary orbital character of the bands.
In Region A, the valence band is primarily derived from N p states, while the conduction band is dominated by Ga s and Ga p states, with strong Rh d-state contributions at the interface that induce band bending in both the conduction and valence bands. In contrast, Region B without nanocluster coverage exhibits well-defined surface states localized near the valence and conduction band edges.

Overall, the region-resolved PLDOS analysis provides a direct spatial picture of how nanocluster adsorption restructures the electronic landscape of GaN-based photocatalysts. Nanocluster-covered regions promote charge injection through localized band bending and interfacial hybridization, while adjacent uncovered regions \hl{preserve surface states that may participate in adsorbate interactions and water activation}. Together, these spatially separated electronic structure responses establish a nanoscale lateral partitioning of photocatalytic functions, providing the physical foundation for the region-resolved interface descriptors introduced in the following section.

\subsection{Region-resolved interface descriptors and interface score construction}
\label{sec:interface_score}

\begin{figure*}
    \centering
    \includegraphics[scale=1]{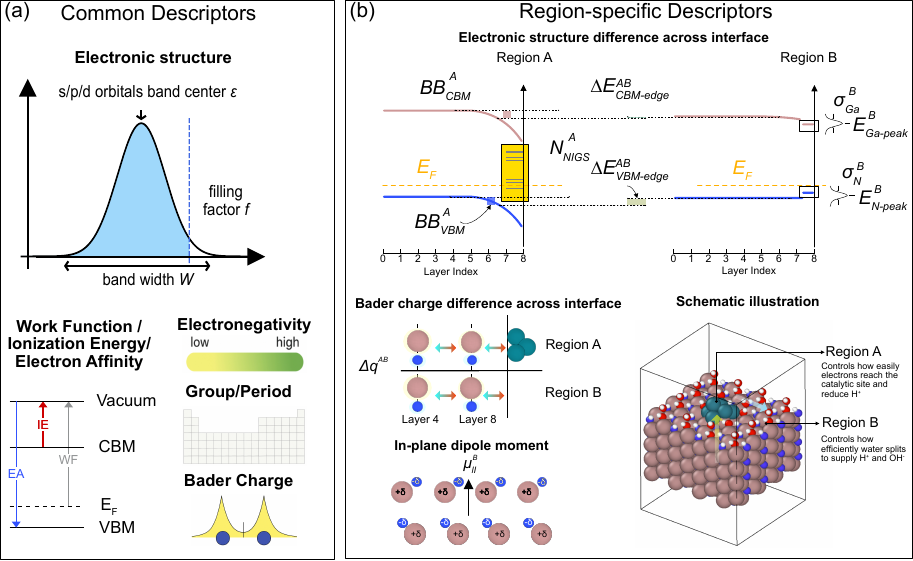}
    \caption{(a) Global descriptors capture composition-averaged electronic and chemical properties, including orbital band centers and widths, filling factors, work function, ionization energy, electron affinity, electronegativity, and Bader charge (see Table \ref{tab:global}). (b) Region-resolved descriptors distinguish the nanocluster-covered region (A) from the uncovered surface (B), quantifying band bending, nanocluster-induced gap states, surface-state peak energies and widths, in-plane dipole moments, lateral band-edge offsets, and interfacial charge redistribution (see Table \ref{tab:iface}). The schematic summarizes the spatial partitioning and functional roles of these descriptors.}
    \label{fig:new_descriptors}
\end{figure*}

Traditional frameworks, most notably the d-band theory for transition-metal catalysis \cite{hammer2000, norskov2011}, successfully rationalize adsorption trends on metallic surfaces through composition-averaged electronic metrics. Chemical descriptors such as group/period position, Bader charge, work function, ionization energy, and electron affinity have therefore been widely employed in regression models to correlate with hydrogen-adsorption energetics in accordance with the Sabatier principle \cite{che2013, medford2015, norskov2005, shirani2023, chen2024, xu2024}. While such global descriptors [see Fig.~\ref{fig:new_descriptors}(a) and Table~\ref{tab:global}] capture broad thermodynamic trends, they obscure the spatially localized interfacial physics governing charge separation and charge utilization that is essential to photocatalytic performance.

Photocatalysis involves far more than the energetics of a single reaction step. Upon illumination, charge carriers are photogenerated in the semiconductor, injected toward the surface, and subsequently drive multi-step redox reactions at spatially distinct sites. The efficiency of this process therefore depends not only on adsorption energetics but also on how effectively electrons and holes are separated, transported, and utilized across heterogeneous interfaces.  These multiscale processes are rarely captured by conventional high-throughput descriptors. Our goal is to bridge this gap by establishing spatially resolved, region-specific descriptors that link electronic structure to interfacial charge dynamics, thereby enabling a more device-oriented design strategy for semiconductor–cocatalyst systems [see Fig.~\ref{fig:new_descriptors}(b) and Table~\ref{tab:iface}]. Interfaces with similar global adsorption energetics may nevertheless exhibit vastly different photocatalytic efficiencies due to differences in local band bending, carrier trapping, and lateral charge separation, effects that are invisible to spatially averaged descriptors.

The proposed framework distinguishes two complementary regions in the GaN-based heterostructure [See Fig.~\ref{fig:new_descriptors}(b) and Table~\ref{tab:iface}]: Region A, directly beneath the nanocluster, governs electron injection, rectification, and hydrogen reduction; 
Region B, the uncovered GaN surface, \hl{is associated with adsorbate interactions and possible water-activation pathways}.
For Region A, we define descriptors that quantify nanocluster-induced perturbations of the host semiconductor: the conduction- and valence-band bending magnitudes ($BB_{\mathrm{CBM}}^\mathrm{A}$, $BB_{\mathrm{VBM}}^\mathrm{A}$), which reflect the built-in interfacial electric field; the nanocluster-induced gap-state density $N_{\mathrm{NIGS}}^\mathrm{A}$, which captures metal-induced hybridization and carrier trapping; and the surface-to-bulk charge redistribution $\Delta q^{A}$. For Region B, descriptors emphasize surface reactivity and electrostatics: the surface-state peak position $E^\mathrm{B}_{\mathrm{peak}}$, spectral width $\sigma^\mathrm{B}$, alongside the surface charge redistribution $\Delta q^{B}$ and in-plane dipole moment $\mu^\mathrm{B}_{\|}$. To capture coupling between the two regions, we further evaluate lateral Bader charge differences $\Delta q^\mathrm{AB}$ and lateral band-edge offsets $\Delta E_{\mathrm{edge}}^\mathrm{AB}$, which quantify charge redistribution and potential steps across the A–B junction. The detailed definitions, extraction procedures, and normalization of these region-resolved descriptors are provided in Appendix~\ref{sec:ap_interface_descriptors}.

Designing efficient semiconductor–nanocluster photocatalysts requires balancing charge injection, surface activation, and interfacial coupling within a unified quantitative framework. Rather than predicting a single optimal interface, our goal is to construct a physically interpretable metric that enables comparative ranking and mechanistic analysis across heterogeneous interfaces.
To this end, we introduce a physics-informed Interface Score (IS) that integrates the region-resolved descriptors developed in Secs.~\ref{sec:PLDOS}–\ref{sec:interface_score} into a unified metric of interfacial efficiency. The IS comprises three contributions corresponding to the physical mechanisms discussed above: Injection ($I_\mathrm{inj}$, Region A), Activation ($I_\mathrm{act}$, Region B), and Coupling ($I_\mathrm{coupl}$, Region A $\leftrightarrow$ B). The mathematical definitions of these sub-scores are provided in Appendix~\ref{sec:ap_IS_score}. These components are combined as
\begin{equation}
\mathrm{IS} = w_A I_\mathrm{inj} + w_B I_\mathrm{act} + w_{AB} I_\mathrm{coupl},
\end{equation}
where $w_A$, $w_B$, and $w_{AB}$ are the relative importance of the three physical processes. 

To examine how the choice of weighting influences interface ranking, we performed an illustrative calibration of the IS parameters using a minimal empirical constraint, namely that at least one Rh-containing interface attains the highest rank among all candidate systems. This constraint reflects the limited but well-established experimental observation that Rh-based cocatalysts perform favorably in III-nitride photocatalytic systems. Under this minimal supervision, both the substrate identity and the internal descriptor weights were allowed to vary freely. The resulting rankings for each nanocluster-substrate combination, exemplified in Table~\ref{tab:top10} (Appendix~\ref{sec:ap_IS_score_cali_example}), are not intended as a definitive optimization of the weighting scheme. Instead, they demonstrate how the Interface Score integrates region-resolved interface descriptors (Table~\ref{tab:iface}) into a physically interpretable comparative metric. As broader experimental benchmarks become available, the same framework can be systematically recalibrated to provide more quantitative guidance. Further details of the calibration procedure, robustness analysis, and representative parameterizations are provided in Appendix~\ref{sec:ap_IS_score_cali_example}.

Collectively, these region-resolved descriptors transform the heterogeneous semiconductor–nanocluster interface into a physically structured, machine-readable representation, forming a direct foundation for data-driven feature attribution and interpretability analysis in the following section.

\begin{table*}[t]
\caption{\label{tab:global}Global descriptors with their units.}
\begin{ruledtabular}
\begin{tabular}{ll}
\noalign{\smallskip}
\textbf{Global descriptors} & \textbf{unit} \\
\hline
\noalign{\smallskip}
surface layer Ga Bader charge & e \\
surface layer N Bader charge & e \\
nanocluster element Bader charge & e \\
H Bader charge & e \\
\noalign{\smallskip}
nanocluster contributions within bands $R^{X}_\mathrm{VBM}$, $R^{X}_\mathrm{GAP}$, $R^{X}_\mathrm{CBM}$ (Sec.~\ref{sec:ap_band_ratios}) & \\
H contributions within bands $R^\mathrm{H}_\mathrm{VBM}$, $R^\mathrm{H}_\mathrm{GAP}$, $R^\mathrm{H}_\mathrm{CBM}$ (Sec.~\ref{sec:ap_band_ratios}) & \\
nanocluster orbital contributions in the VBM $R_\mathrm{VBM}^\mathrm{s}$, $R_\mathrm{VBM}^\mathrm{p}$, $R_\mathrm{VBM}^\mathrm{d}$ (Sec.~\ref{sec:ap_orbital_vbm_ratios}) & \\
nanocluster s/p/d center/bandwidth/edge (Sec.~\ref{sec:ap_band_properties}) & \\
H s center/bandwidth (Sec.~\ref{sec:ap_band_properties}) & \\
H deep-s center/bandwidth (Sec.~\ref{sec:ap_H_deep_s}) & \\
\noalign{\smallskip}
Fermi Level $E_\mathbf{F}$ & eV \\
Work Function & eV \\
Group & \\
Period & \\
\noalign{\smallskip}
\end{tabular}
\end{ruledtabular}
\end{table*}

\begin{table*}[t]
\caption{\label{tab:iface}Region-resolved electronic descriptors and their physical roles in GaN–nanocluster photocatalysts, illustrated schematically in Fig.~\ref{fig:new_descriptors}(b).}
\begin{ruledtabular}
\begin{tabular}{lll}
\noalign{\smallskip}
\textbf{Axis} & \textbf{Typical descriptors} & \textbf{Physical role} \\
\hline
\noalign{\medskip}
\parbox[c]{3.5cm}{\raggedright \textbf{Activation / Volmer} \\ (Region B)} &
\parbox[c]{4.5cm}{\raggedright $E_{\text{peak}}^{\mathrm{B}}$,  $\sigma^{\mathrm{B}}$, $\mu_{\|}^{\mathrm{B}}$, $\Delta q^{\mathrm{B}}$ \\ (Sec.~\ref{sec:ap_B_descrip})} &
\parbox[c]{6cm}{\raggedright Controls how efficiently water splits to supply H$^{+}$ and OH$^{-}$.} \\
\noalign{\medskip}
\parbox[c]{3.5cm}{\raggedright \textbf{Injection / Reduction} \\ (Region A)} &
\parbox[c]{4.5cm}{\raggedright $BB_{\mathrm{CBM}}^{\mathrm{A}}$, $BB_{\mathrm{VBM}}^{\mathrm{A}}$, $N_{\text{NIGS}}^{\mathrm{A}}$, $\Delta q^{\mathrm{A}}$ \\ (Sec.~\ref{sec:ap_A_descrip})} &
\parbox[c]{6cm}{\raggedright Controls how easily electrons reach the catalytic site and reduce H$^{+}$.} \\
\noalign{\medskip}
\parbox[c]{3.5cm}{\raggedright \textbf{Interfacial coupling} \\ (Between Region A \& B)} &
\parbox[c]{4.5cm}{\raggedright $\Delta E_{\text{band-edge}}^\mathrm{AB}$, $\Delta q^{\mathrm{AB}}$ \\ (Sec.~\ref{sec:ap_iface_descrip})} &
\parbox[c]{6cm}{\raggedright Quantifies the lateral potential step and charge redistribution bridging the two regions.} \\
\noalign{\smallskip}
\end{tabular}
\end{ruledtabular}
\end{table*}

\subsection{Data-driven feature attribution and physical interpretability}
\label{sec:ML}

To identify the dominant interfacial mechanisms governing hydrogen adsorption, we employ supervised regression models primarily as an interpretability tool, rather than as black-box predictors.
\hl{
The hydrogen adsorption energy ($E_{\mathrm{H}}$) is explicitly defined as:}
\begin{equation}
    E_{\mathrm{H}} = E_{\mathrm{slab}+\mathrm{NC}+\mathrm{H}} - E_{\mathrm{slab}+\mathrm{NC}} - \frac{1}{2} E_{\mathrm{H}_2}
\end{equation}
\hl{where $E_{\mathrm{slab}+\mathrm{NC}+\mathrm{H}}$, $E_{\mathrm{slab}+\mathrm{NC}}$, and $E_{\mathrm{H}_2}$ are the DFT-calculated total energies of the hydrogen-adsorbed system, the bare semiconductor-nanocluster interface, and an isolated hydrogen molecule, respectively.
} 
\hl{It should be noted that while our regression models map electronic descriptors to the ground-state adsorption energy ($E_{\mathrm{H}}$), practical catalytic rates are governed by the Gibbs free energy ($\Delta G_{\mathrm{H}}$). Because calculating exact vibrational frequencies across all heterogeneous interfaces is computationally demanding, and because thermodynamic corrections for H-adsorption are often approximated as a constant shift in large-scale machine-learning studies }\cite{xu2024}, \hl{we focus here on $E_{\mathrm{H}}$ to extract baseline feature attributions.}

Here, two complementary descriptor sets are considered: (i) global chemical descriptors based on band-center and bandwidth metrics, group/period identifiers, and Bader charges; and (ii) the region-resolved interface descriptors introduced in Section~\ref{sec:interface_score}, which explicitly encode spatially differentiated interfacial responses. By comparing these representations within a unified modeling framework, we assess not only predictive performance but, more importantly, the extent to which physically motivated, region-resolved features enable mechanistic insight. The full descriptor sets are summarized in Tables~\ref{tab:global} and~\ref{tab:iface}, with modeling pipelines, cross-validation procedures, and hyperparameter optimization detailed in Sec.~\ref{sec:ap_ml_method}.

\begin{figure*}
\centering
\includegraphics[scale=1]{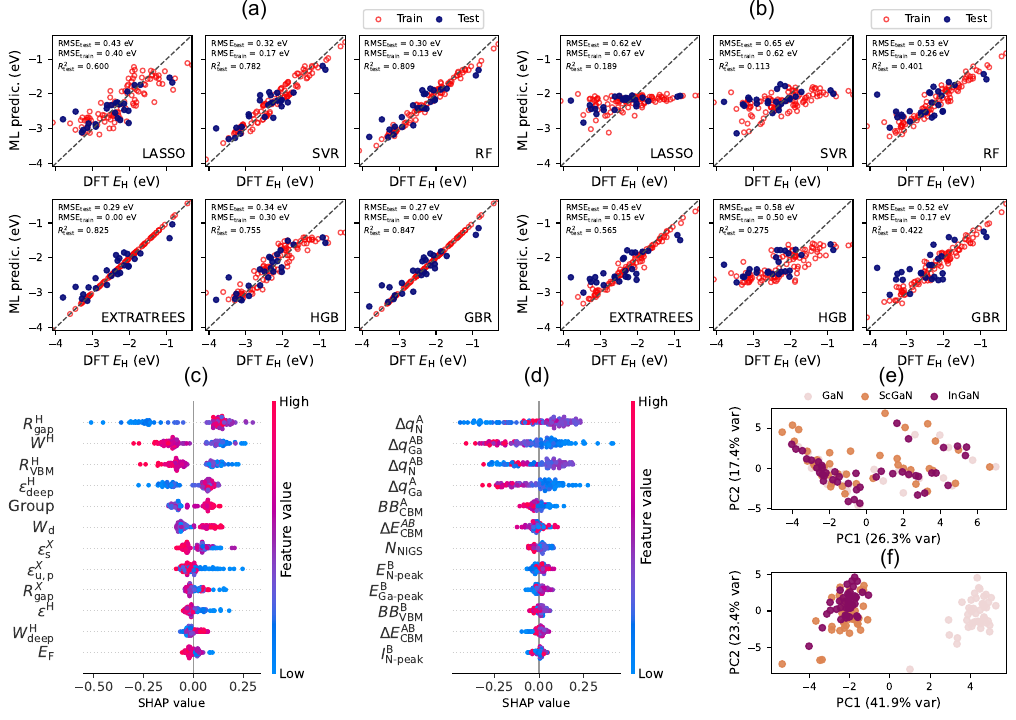}
\caption{(a,b) Parity plots comparing machine-learning predictions with DFT-calculated hydrogen adsorption energies $E_\mathrm{H}$ using (a) global descriptors and (b) region-resolved interface descriptors across multiple regression algorithms. Red and blue markers denote training and test sets, respectively; dashed lines indicate perfect agreement. (c,d) SHAP analysis of the best-performing models for (c) the global-descriptor set and (d) the region-resolved descriptor set, highlighting the most influential features and their directional contributions to $E_\mathrm{H}$. Feature color denotes relative feature magnitude. (e,f) Principal-component analysis (PCA) projections of (e) the global descriptors  and (f) region-resolved descriptors, colored by substrate (GaN, InGaN, ScGaN), illustrating the enhanced substrate-specific separation captured by spatially resolved interface descriptors.}
\label{fig:ML}
\end{figure*}

Figs.~\ref{fig:ML}(a–b) summarize the regression performance across several models. In the following discussion, we focus on feature attribution rather than absolute prediction accuracy, as the primary objective is to uncover the physical drivers of adsorption energetics. Models based on global descriptors consistently achieve higher $R^2$ values, reflecting the strong predictive signal contained in composition-averaged band alignment and chemical identity. \hl{Because $E_{\mathrm{H}}$ is fundamentally a macroscopic thermodynamic scalar derived from the total energy difference of the supercell, it naturally exhibits a tighter mathematical correlation with these spatially averaged properties.} In contrast, models trained solely on region-resolved interface descriptors attain moderate but nontrivial predictive performance (e.g., $R^2 \approx 0.57$ with ExtraTrees). \hl{This relative reduction in predictive accuracy is expected, as predicting a single global energy metric from decoupled local electrostatic fields inherently introduces variance. Nevertheless, this performance  } indicates that spatially localized interfacial features encode meaningful catalytic trends when coupled with sufficiently expressive learners.

SHapley Additive exPlanations (SHAP) provide direct insight into how individual electronic descriptors influence the predicted hydrogen adsorption energy $E_\mathrm{H}$. In this representation, negative SHAP values correspond to contributions that lower $E_\mathrm{H}$ (stabilizing adsorption), while positive values indicate contributions that weaken binding.  The color scale encodes the magnitude of each descriptor (red: high values; blue: low values), enabling direction-dependent physical trends to be identified.  For the global descriptor model [Fig.~\ref{fig:ML}(c)], features such as $R_{\mathrm{gap}}^\mathrm{H}$, $W^\mathrm{H}$, $R_{\mathrm{VBM}}^\mathrm{H}$, and $\varepsilon_{\mathrm{deep}}^{\mathrm{H}}$ dominate, confirming that HER thermodynamics are largely controlled by the energetic alignment and filling of H-derived states relative to the host valence manifold. 

By contrast, SHAP analysis of the region-resolved interface model [Fig.~\ref{fig:ML}(d)] highlights descriptors associated with layer- and region-specific charge redistribution, including
$\Delta q_\text{N}^\mathrm{A}$,
$\Delta q_{\mathrm{Ga}}^\mathrm{AB}$, 
$\Delta q_{\mathrm{N}}^\mathrm{AB}$, 
and $\Delta q_{\text{Ga}}^\mathrm{A}$.
High values of Ga-related charge-transfer descriptors, particularly $\Delta q_{\mathrm{Ga}}^\mathrm{AB}$, are predominantly associated with negative SHAP values, indicating that enhanced electron transfer across the lateral A-B junction stabilizes adsorbed hydrogen. In contrast, low values of N-related surface charge descriptors, such as  $\Delta q_\text{N}^\mathrm{A}$ and $\Delta q_{\mathrm{N}}^\mathrm{AB}$, preferentially contribute to negative SHAP values, revealing that excessive electron accumulation on surface N sites destabilizes H adsorption by competing for charge. Together, these trends indicate a non-monotonic, site-specific role of interfacial charge redistribution: effective interfaces promote electron delivery through Ga sites while suppressing competing charge localization on N-derived surface states. 
This interpretation is consistent with the PLDOS-resolved electronic picture discussed in Section~\ref{sec:PLDOS}.

Dimensionality-reduction analyses further elucidate the qualitative differences between the two descriptor sets. Principal Component Analysis (PCA) applied to global descriptors [Fig.~\ref{fig:ML}(e)] yields diffuse, overlapping clusters, reflecting the dominance of broad chemical similarity with limited spatial specificity. In contrast, PCA embeddings of the region-resolved descriptors [Fig.~\ref{fig:ML}(f)] exhibit clear separation by substrate composition (GaN, InGaN, and ScGaN). This clustering demonstrates that the collective band bending, charge redistribution, and surface-state characteristics encoded in the region descriptors are strongly substrate dependent. This behavior underscores the central role of substrate-controlled electrostatics in defining the interfacial operating regime, which connects to the design principle that the substrate establishes the baseline electrostatic landscape within which the nanocluster chemistry acts as a local tuner.

Taken together, these results show that while global descriptors provide higher raw predictive accuracy for hydrogen adsorption energies, region-resolved descriptors offer greater mechanistic interpretability by explicitly encoding spatially dependent interfacial physics. Multi-tiered hybrid strategies can thus be viewed as complementary workflows, in which global descriptors are used for rapid screening of adsorption-dominated trends associated with nanocluster-derived electronic structure within a given substrate platform, while region-resolved descriptors or the Interface Score (IS) introduced in Section~\ref{sec:interface_score} are employed to refine and prioritize electrostatics- and transport-limited behavior upon subsequent interface and surface optimization in selected nanocluster–semiconductor systems.

\section{Discussion}
\label{sec:Discussion}
This work \hl{provides a region-resolved analysis of how nanocluster cocatalysts modify III-nitride surfaces by reshaping interfacial electronic structure and charge redistribution.} By combining region-resolved PLDOS analysis with physically motivated electronic descriptors and interpretable machine-learning models, we move beyond treating the semiconductor-nanocluster interface as electronically homogeneous. Instead, the interface emerges as a laterally heterogeneous electronic system in which spatially distinct regions play complementary roles in charge separation, transport, and surface chemistry.

Global electronic descriptors, dominated by hydrogen band-center positions, bandwidths, and projected density-of-states features, achieve the highest predictive accuracy for hydrogen adsorption energies. This observation reinforces the well-established understanding that adsorption thermodynamics are largely governed by energetic alignment between H-derived states and the host electronic structure \cite{hammer2000, norskov2011, medford2015, norskov2005}. From a design standpoint, this confirms that alloy composition and nanocluster chemistry remain first-order parameters for tuning overall reactivity, consistent with conventional band-alignment and d-band based arguments.

At the same time, global descriptors alone obscure how photocatalytic efficiency emerges at spatially heterogeneous interfaces. \hl{Thus, interfaces with comparable hydrogen adsorption energies may still differ substantially in photocatalytic behavior if they exhibit different degrees of band bending, carrier trapping, or lateral charge separation.}  The region-resolved descriptors introduced here expose an additional and essential layer of physics. PLDOS analysis shows that nanocluster adsorption induces pronounced band bending and nanocluster-induced gap states beneath the cocatalyst (Region A), while uncovered III-nitride surface regions (Region B) retain well-defined surface states and intrinsic in-plane dipoles. These features naturally map onto distinct photocatalytic functions: 
Region A \hl{appears more closely associated with} charge injection and electron accumulation at the cocatalyst, whereas Region B \hl{may contribute to} electronic conditions relevant to water activation and proton-related surface chemistry.
In this sense, semiconductor-nanocluster interfaces operate as nanoscale lateral heterojunctions rather than uniform catalytic surfaces, directly reflecting the first design \hl{insight} outlined in the Introduction, which emphasizes the central role of lateral heterogeneity in semiconductor-nanocluster interfaces.

At a quantitative level, machine-learning analyses based on the region-resolved descriptors further clarify how interfacial electronic structure influences charge separation and transport. While global descriptors tied to adsorption energetics or averaged band alignment can correlate strongly with hydrogen binding, they necessarily compress the interface into a spatially homogeneous picture and do not encode local electrostatic responses such as band bending, surface-state trapping, or nanocluster-induced gap states. By contrast, the region-resolved descriptors are constructed to capture these local variations in electrostatics and charge redistribution across laterally distinct regions. This perspective reflects the second design \hl{insight}, emphasizing the importance of designing interfacial electrostatics beyond adsorption energetics or averaged band alignment. Moreover, in low-dimensional projections of the region-resolved descriptor space, interfaces naturally cluster by substrate (GaN, InGaN, and ScGaN), indicating that substrate composition sets the dominant interfacial electrostatic regime through its influence on band bending, surface states, and charge redistribution. This observation leads to the third design \hl{insight}, namely that substrate selection defines the baseline electrostatic landscape within which nanocluster chemistry operates.

Together, these results motivate a hierarchical design strategy for semiconductor nanocluster photocatalysts.
The choice of substrate establishes the electrostatic baseline, while nanocluster properties and nanocluster-substrate coupling shape local and lateral band bending, nanocluster-induced gap states, and the resulting charge separation capability. Interface engineering strategies, including alloying, cocatalyst selection, and surface dipole control, can then be used to optimize lateral coupling between Regions A and B, suppress recombination pathways, and promote directional carrier flow, directly \hl{consistent with the physical insights} outlined in this work.

The Interface Score (IS) integrates these region-resolved physical effects into a comparative metric that enables direct comparison of heterogeneous interfaces while retaining clear physical interpretability. Rather than serving as a predictor of a unique optimal configuration, the IS provides a flexible framework for comparing relative interfacial efficiencies and guiding physics-informed screening across complex interface spaces. As experimental benchmarks for photocatalytic hydrogen evolution become available, the same framework can be systematically refined or adapted to different operating conditions. More broadly, the IS offers a natural bridge between first-principles electronic-structure analysis and active-learning or optimization workflows for exploring large interface design spaces.

\section{Conclusion}
\label{sec:conclusion}

We present a physics-guided framework for understanding and designing semiconductor–nanocluster photocatalytic interfaces by explicitly resolving their lateral electronic heterogeneity. Using a systematic \textit{ab initio} dataset of GaN-, InGaN-, and ScGaN-based nanocluster interfaces, we show that these systems \hl{can be described} not as uniform catalytic surfaces, but as laterally heterogeneous interfaces in which spatially distinct regions exhibit different electronic responses. Region-resolved PLDOS analysis and physically motivated electronic descriptors indicate that nanocluster-covered regions are associated with charge injection, local band bending, and nanocluster-induced gap states, while adjacent uncovered nitride regions retain surface-localized states and electrostatic features \hl{that may influence adsorbate interactions and surface activation.} \hl{Interpretable machine-learning analyses} are used to assess the correlation between electronic descriptors and single-H adsorption energetics, showing that global descriptors capture broad adsorption-energy trends while region-resolved descriptors provide complementary insight into local interfacial electrostatics and charge redistribution.
\hl{In addition, the Interface Score provides a physics-informed way to combine region-resolved interface descriptors into a single comparative metric, which may help compare and screen heterogeneous model interfaces on a common physical basis.}
Although demonstrated for III-nitride systems, the underlying concepts are broadly applicable to heterogeneous semiconductor interfaces where lateral electronic structure governs energy-conversion processes.

\section*{Data and Code Availability}
\label{sec:ap_avail}
An interactive visualization of the GaN-nanocluster interface database, including structural files, electronic structure plots, and descriptor values, is available at
\url{https://gan-nanocluster-database-ml.vercel.app/}.
The full underlying datasets and analysis scripts supporting this study will be made publicly available upon acceptance.

\section*{Acknowledgements}
S. Yuan thanks Zhanghao Zhouyin for helpful discussions and constructive comments on the manuscript, particularly regarding the machine-learning methodology. S. Yuan also thanks Yiqing Chen for valuable discussions on computational theories and studies of photocatalysis. S. Yuan and H. Guo gratefully acknowledge financial support from the Global Hydrogen Production Technologies (HyPT) Center, NSERC of Canada and FRQNT of Qu\'ebec.  Computational resources were provided by the Digital Research Alliance of Canada and Calcul Qu\'ebec, which made this work possible. Z. Mi gratefully acknowledges support from the National Science Foundation under grant no. 2330525 as part of the Global Hydrogen Production Technologies (HyPT) Center, and the United States Army Research Office Award W911NF2110337. G. Metha and G. Andersson acknowledge support from CSIRO, Australia’s national science agency, which enables Australian researchers to participate in the HyPT Center with partners in the United States, the United Kingdom, and Canada as part of the NSF Global Centers program, supported by the Science and Industry Endowment Fund.


\appendix

\section{Global Descriptors}
\label{sec:ap_global_descritor}
\setcounter{figure}{0}
\renewcommand{\thefigure}{A\arabic{figure}}
\setcounter{table}{0}
\renewcommand{\thetable}{A\arabic{table}}

\subsection{Calculation of  $R_{\mathrm{VBM}}$, $R_{\mathrm{Gap}}$, and $R_{\mathrm{CBM}}$}
\label{sec:ap_band_ratios}

To analyze the projected density of states (PDOS) contributions from various atomic species, we calculated the relative integrated PDOS in three energy regions: the valence band maximum (VBM), the band gap (Gap), and the conduction band minimum (CBM). In this work, the band-edge markers are defined by surface states: we take \hl{$E^\mathrm{eff}_{\mathrm{VBM}}$} and \hl{$E^\mathrm{eff}_{\mathrm{CBM}}$} as the energies of the N- and Ga-derived surface-state peaks, respectively, and use these as the integration boundaries. For a given element $X$, the ratios 
\begin{equation}
\begin{aligned}
R_{\mathrm{VBM}}^{X} &= \frac{\int_{E_{\mathrm{min}}}^{E^\mathrm{eff}_{\mathrm{VBM}}} \mathrm{PDOS}^{X}(\varepsilon)\, d\varepsilon}{\int_{E_{\mathrm{min}}}^{E_{\mathrm{max}}} \mathrm{PDOS}^{X}(\varepsilon)\, d\varepsilon}, 
\\
R_{\mathrm{Gap}}^{X} &= \frac{\int_{E^\mathrm{eff}_{\mathrm{VBM}}}^{E^\mathrm{eff}_{\mathrm{CBM}}} \mathrm{PDOS}^{X}(\varepsilon)\, d\varepsilon}{\int_{E_{\mathrm{min}}}^{E^{\mathrm{max}}} \mathrm{PDOS}^{X}(\varepsilon)\, d\varepsilon}, 
\\
R_{\mathrm{CBM}}^{X} &= \frac{\int_{E^\mathrm{eff}_{\mathrm{CBM}}}^{E_{\mathrm{max}}} \mathrm{PDOS}^{X}(\varepsilon)\, d\varepsilon}{\int_{E_{\mathrm{min}}}^{E_{\mathrm{max}}} \mathrm{PDOS}^{X}(\varepsilon)\, d\varepsilon},
\end{aligned}
\end{equation}
where $E_{\mathrm{min}} = -10$ eV and $E_{\mathrm{max}} = +10$ eV relative to the Fermi level. The positions of \hl{$E^\mathrm{eff}_{\mathrm{VBM}}$} and \hl{$E^\mathrm{eff}_{\mathrm{CBM}}$} were determined by identifying the local PDOS peaks corresponding to surface states of Ga and N, respectively. The integrals were evaluated numerically using the trapezoidal rule. The same procedure was applied to hydrogen atoms to obtain $R_{\mathrm{VBM}}^\mathrm{H}$, $R_{\mathrm{Gap}}^\mathrm{H}$, and $R_{\mathrm{CBM}}^\mathrm{H}$.

The total PDOS integral for each species, $\int_{E_{\mathrm{min}}}^{E_{\mathrm{max}}} \mathrm{PDOS}^{X}(E)\, dE$, was used to normalize the contributions in each energy region. This approach enabled a consistent comparison of relative PDOS contributions across different atomic species and energy ranges.

\begin{figure*}
    \centering
    \includegraphics[scale=0.6]{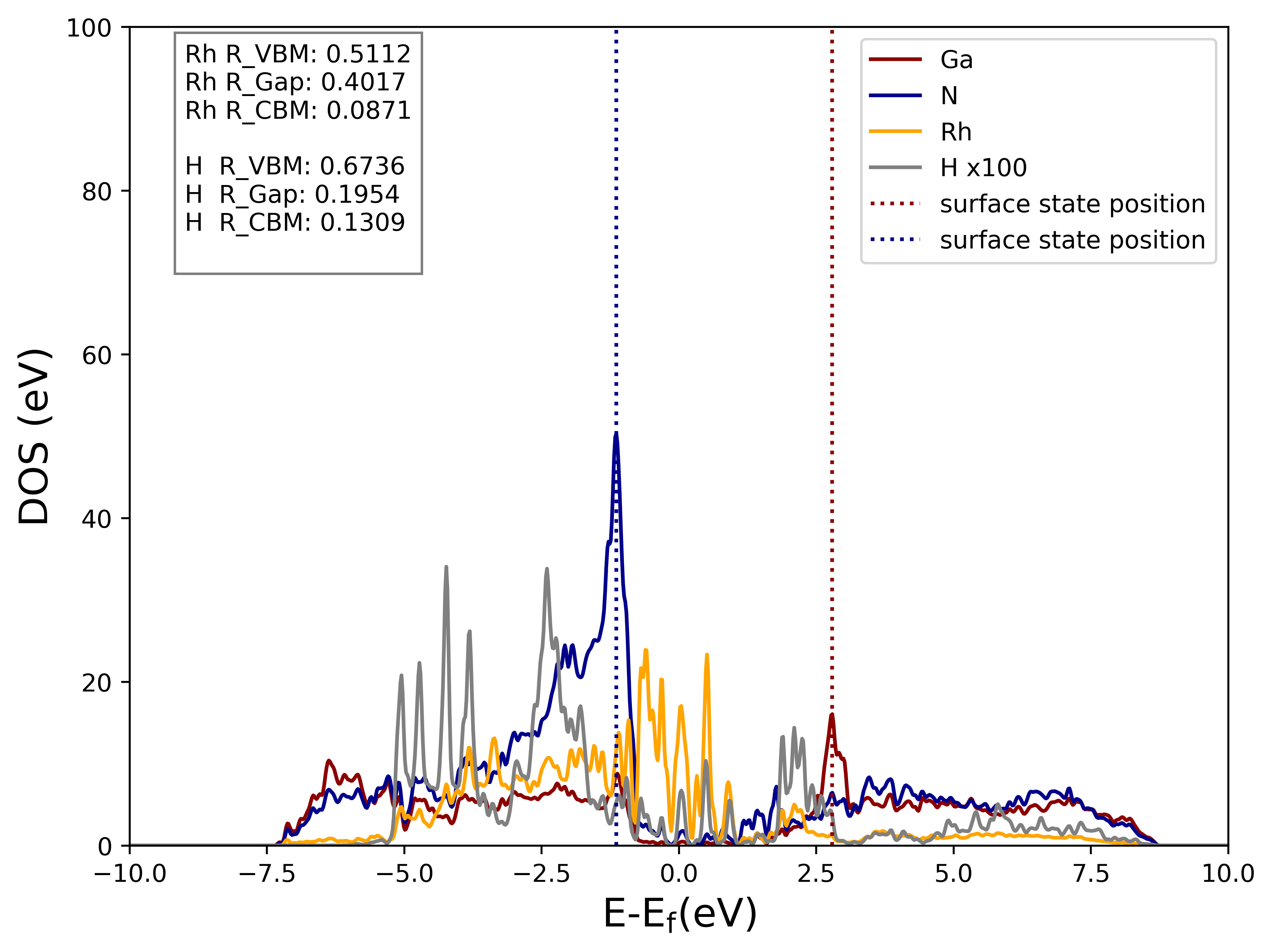}
    \caption{Element-resolved PDOS of Rh and H  on the GaN slab, highlighting the surface state positions ($E_{\mathrm{VBM}}$ and $E_{\mathrm{CBM}}$) used to define integration windows. The computed ratios for Rh are: $R_{\mathrm{VBM}} = 0.5112$, $R_{\mathrm{Gap}} = 0.4017$, and $R_{\mathrm{CBM}} = 0.0871$. For H they are: $R_{\mathrm{VBM}} = 0.6736$, $R_{\mathrm{Gap}} = 0.1954$, and $R_{\mathrm{CBM}} = 0.1309$.}

    \label{fig:Rh_ratio_extract}
\end{figure*}

As an illustrative example, Fig.~\ref{fig:Rh_ratio_extract} shows the element-resolved PDOS of rhodium (Rh) and hydrogen (H) on the GaN surface, with vertical lines marking the identified surface state positions $E_{\mathrm{VBM}}$ and $E_{\mathrm{CBM}}$. The corresponding ratios for Rh are:
\[
R_{\mathrm{VBM}}^{\mathrm{Rh}} = 0.5112, \quad
R_{\mathrm{Gap}}^{\mathrm{Rh}} = 0.4017, \quad
R_{\mathrm{CBM}}^{\mathrm{Rh}} = 0.0871
\]
while for H they are:
\[
R_{\mathrm{VBM}}^{\mathrm{H}} = 0.6736, \quad
R_{\mathrm{Gap}}^{\mathrm{H}} = 0.1954, \quad
R_{\mathrm{CBM}}^{\mathrm{H}} = 0.1309.
\]
These results demonstrate the relative contributions of Rh and H to different energy regions, providing insight into their potential roles in photocatalytic processes. This analysis was repeated systematically for all relevant elements in our dataset.

\subsection{Calculation of $R^{\mathrm{s}}_{\mathrm{VBM}}$, $R^{\mathrm{p}}_{\mathrm{VBM}}$, and $R^{\mathrm{d}}_{\mathrm{VBM}}$}
\label{sec:ap_orbital_vbm_ratios}

\begin{equation}
I_{\mathrm{VBM}}^{{X},\mathrm{l}} \;=\; \int_{E_{\min}}^{E_{\mathrm{VBM}}} \mathrm{PDOS}^{{X},\mathrm{l}}(E)\, dE,
\qquad \mathrm{l} \in \{\mathrm{s,p,d}\},
\end{equation}
using the trapezoidal rule on the nonuniform energy grid. The VBM orbital fractions were then obtained by normalizing to the total $s$+$p$+$d$ weight within the same interval,
\begin{equation}
R_{\mathrm{VBM}}^{{X},\mathrm{l}}
\;=\;
\frac{I_{\mathrm{VBM}}^{{X},\mathrm{l}}}
{I_{\mathrm{VBM}}^{{X},\mathrm{s}}+I_{\mathrm{VBM}}^{{X},\mathrm{p}}+I_{\mathrm{VBM}}^{{X},\mathrm{d}}},
\qquad \mathrm{l} \in \{\mathrm{s,p,d}\}.
\end{equation}
This definition isolates the relative $s$/$p$/$d$ composition right at the valence-band edge, independent of the absolute DOS magnitude of element $X$. For completeness, the same integration procedure also yields $I_{\mathrm{Gap}}^{{X},\mathrm{l}}$ and $I_{\mathrm{CBM}}^{{X},\mathrm{l}}$ over $[E_{\mathrm{VBM}},E_{\mathrm{CBM}}]$ and $[E_{\mathrm{CBM}},E_{\max}]$, respectively.

\subsection{Calculation of s/p/d-band properties}
\label{sec:ap_band_properties}
To further analyze the electronic structure of the system, we evaluated the d-band center ($\varepsilon_\mathrm{d}$), the d-band width ($W_\mathrm{d}$), and the upper edge of the d-band ($\varepsilon_\mathrm{u}$) using the projected density of states (PDOS) resolved in the energy range $[E_{\mathrm{min}}, E_{\mathrm{max}}]$. The d-band center was computed as the first moment of the PDOS:

\begin{equation}
\varepsilon_\mathrm{d} = \frac{\int_{E_{\mathrm{min}}}^{E_{\mathrm{max}}} E \, \mathrm{PDOS}(E) \, dE}{\int_{E_{\mathrm{min}}}^{E_{\mathrm{max}}} \mathrm{PDOS}(E) \, dE}.
\end{equation}

The d-band width was defined as the square root of \hl{the central second moment}:

\begin{equation}
W_\mathrm{d} = \sqrt{\frac{\int_{E_{\mathrm{min}}}^{E_{\mathrm{max}}} (E-\varepsilon_{\mathrm{d}})^2 \, \mathrm{PDOS}(E) \, dE}{\int_{E_{\mathrm{min}}}^{E_{\mathrm{max}}} \mathrm{PDOS}(E) \, dE}}.
\end{equation}

The upper edge of the d-band, $\varepsilon_\mathrm{u}$, was determined from the maximum of the Hilbert transform of the PDOS:

\begin{equation}
\varepsilon_\mathrm{u} = \mathrm{argmax}\left[ \mathcal{H}\left\{\mathrm{PDOS}(E)\right\} \right]
\end{equation}

where the Hilbert transform was evaluated numerically using a small imaginary broadening $\eta$:

\begin{equation} 
\mathcal{H}\left\{\mathrm{PDOS}(E)\right\}\!=\!\frac{1}{\pi} \int_{E_{\mathrm{min}}}^{E_{\mathrm{max}}} \mathrm{PDOS}(E') \frac{E - E'}{(E - E')^2 + \eta^2} dE'.
\end{equation}

In this work, we used $E_{\mathrm{min}} = -10$ eV and $E_{\mathrm{max}} = +10$~eV relative to the Fermi level, and $\eta = 0.01$ eV. All integrals were evaluated numerically using the trapezoidal rule to ensure consistency with the discrete energy grid of our calculations. When the total PDOS integral was negligible (or in the presence of numerical instabilities), default values were assigned to avoid nonphysical results.

In addition to the d-orbital analysis, the same approach was applied to the s and p PDOS projections to obtain their respective centers and widths, enabling a comprehensive comparison across different orbital contributions.

\begin{figure*}
    \includegraphics[scale=0.6]{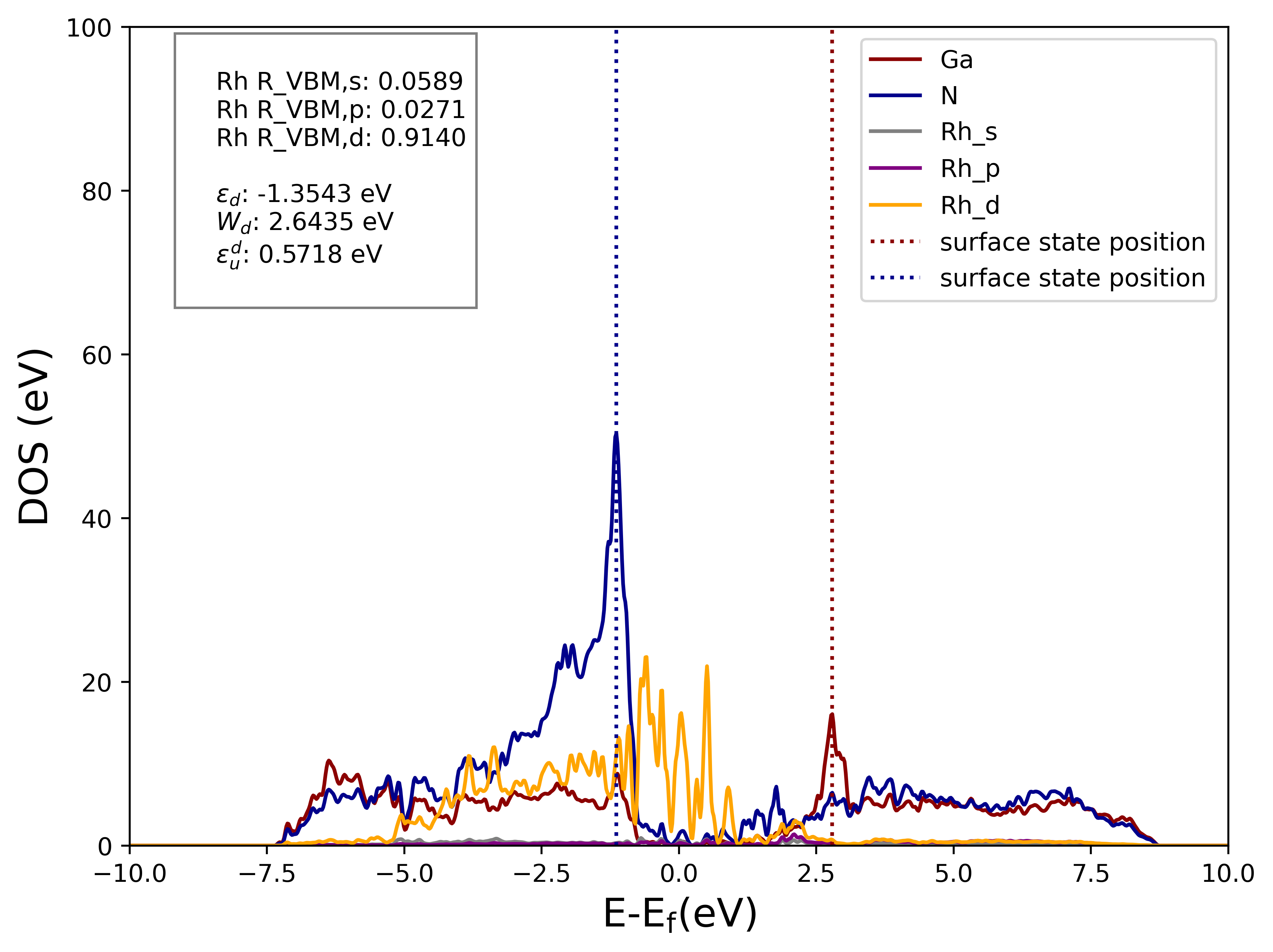}
    \caption{Projected density of states (PDOS) of Ga, N, and Rh ($s$, $p$, $d$ orbitals) on the GaN slab. Vertical dashed lines indicate the identified surface state positions (VBM and CBM). The insets summarize the relative VBM contributions and d-band properties from each Rh orbital.}
    \label{fig:Rh_ratio_extract_orbitals}
\end{figure*}

As an illustrative example, we analyzed a Rh cluster deposited on an InGaN slab. Fig.~\ref{fig:Rh_ratio_extract_orbitals} shows the PDOS contributions from Ga, N, and Rh (resolved into $s$, $p$, and $d$ orbitals). The calculated VBM contributions of Rh are: $R_{\mathrm{VBM},\mathrm{s}} = 0.0589$, $R_{\mathrm{VBM},\mathrm{p}} = 0.0271$, and $R_{\mathrm{VBM},\mathrm{d}} = 0.9140$, indicating that the $\mathrm{d}$ states dominate near the VBM region.

The d-band properties of Rh were determined as:
\begin{equation}
\begin{aligned}
& \varepsilon_\mathrm{d} = -1.3543  \ \mathrm{eV}, \\
& W_\mathrm{d} = 2.6435 \ \mathrm{eV}, \\
&\varepsilon_{\mathrm{u},\mathrm{d}} = 0.5718 \ \mathrm{eV}.
\end{aligned}
\end{equation}

Similarly, for the $s$ and $p$ orbitals, we obtained:
\begin{equation}
\begin{aligned}
&  \varepsilon_\mathrm{s} = -0.3594 \ \mathrm{eV},     \ &\varepsilon_\mathrm{p} = 2.5069 \ \mathrm{eV},                  \\
&  W_\mathrm{s} = 3.9107 \ \mathrm{eV},                \ & W_\mathrm{p} = 4.4485 \ \mathrm{eV},                 \\
&  \varepsilon_\mathrm{u,s} = -0.2632 \ \mathrm{eV},    \  &  \varepsilon_\mathrm{u,p} = 7.5778 \ \mathrm{eV}.                  \\
\end{aligned}
\end{equation}

\subsection{Hydrogen deep-$s$ descriptor}
\label{sec:ap_H_deep_s}
In addition to the $[-10,10]$~eV window used for edge-resolved ratios, we compute a hydrogen-specific descriptor that targets deep, ``semi-core–like’’ $1$s states. Specifically, we integrate the H-s PDOS over $[-70,-15]$~eV (relative to $E_{\mathrm{F}}$),
\begin{equation}
A^{\mathrm{H,s}}_{\mathrm{deep}} \;=\; \int_{-70~\mathrm{eV}}^{-15~\mathrm{eV}} \mathrm{PDOS}^{\mathrm{H,s}}(E)\, dE,
\end{equation}
and report its first and second moments (band center and width),
\begin{equation}
\begin{aligned}
\varepsilon^{\mathrm{H,s}}_{\mathrm{deep}} \;=\;  &
\frac{\int_{-70}^{-15} E\,\mathrm{PDOS}^{\mathrm{H,s}}(E)\, dE}{A^{\mathrm{H,s}}_{\mathrm{deep}}}, 
\qquad \\
W^{\mathrm{H},s}_{\mathrm{deep}} \;=\; &
\sqrt{\frac{\int_{-70}^{-15} E^{2}\,\mathrm{PDOS}^{\mathrm{H},s}(E)\, dE}{A^{\mathrm{H},s}_{\mathrm{deep}}}}.
\end{aligned}
\end{equation}
The $[-70,-15]$~eV range is chosen to exclude the valence manifold while capturing deep H–nanoclusters bonding states. This descriptor is evaluated with the same trapezoidal integration used elsewhere and is reported separately from the $R^{X,\ell}_{\mathrm{VBM}}$ normalization, providing complementary sensitivity to hydrogen’s deep $s$-state character when such semicore contributions influence the observed trends.

\section{Interface Descriptors}
\label{sec:ap_interface_descriptors}
\setcounter{figure}{0}
\renewcommand{\thefigure}{B\arabic{figure}}
\setcounter{table}{0}
\renewcommand{\thetable}{B\arabic{table}}

\subsection{Activation (Region B)}
\label{sec:ap_B_descrip}

\subsubsection{Region B surface states metrics}
\label{sec:ap_B_descrip_surface_states}

\begin{figure*}
    \centering
    \includegraphics[scale=0.6]{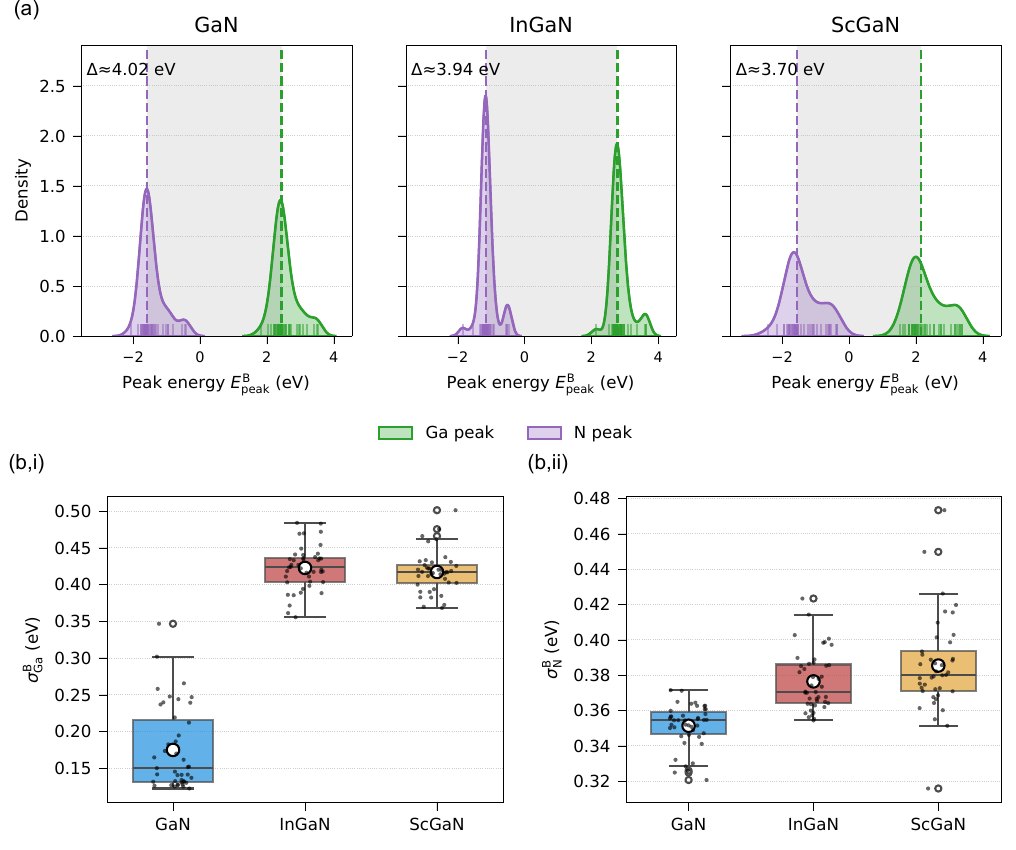}
    \caption{Region B surface-state characteristics.
(a) Distributions of Ga (green) and N (purple) surface-state peak energies ($E^\mathrm{B}_\mathrm{Ga-peak}$ and $E^\mathrm{B}_\mathrm{N-peak}$ ) in Region B for GaN, InGaN, and ScGaN substrates, with dashed lines marking medians and shaded bands indicating the Ga- and N-peak energy separation $\Delta$.
(b,i) and (b,ii) show the corresponding distributions of Ga and N surface-state peak widths ($\sigma^\mathrm{B}_{\mathrm{Ga}}$ and  $\sigma^\mathrm{B}_{\mathrm{N}}$). These results reveal how alloy composition modulates the energetic alignment and spatial broadening of Region B surface states.} 
    \label{fig:surface_states}
\end{figure*}

To characterize the energy position and spectral sharpness of the uncovered Region B surface states, the element-resolved projected local density of states (PLDOS) of the top semiconductor layer (Layer 8) was analyzed.
For each element $X$(Ga or N), the $\mathrm{PLDOS}_{X}(E)$ was aligned to the Fermi level and restricted to the relevant surface-resonance window: $E {\in[-5,0]}$ $\mathrm{eV}$ for N-derived valence states and $E {\in[0,5]}$ $\mathrm{eV}$ for Ga-derived conduction states.

The characteristic peak energy $E^\mathrm{B}_{X\text{-peak}}$ was identified as the energy at which $\operatorname{PDOS}_X(E)$ reaches its maximum, with the corresponding maximum intensity defined as the peak height $I^\mathrm{B}_{X\text{-peak}}$.
A local integration window $\left[E_1, E_2\right]$ around $E^\mathrm{B}_{X\text{-peak}}$ was then determined by expanding toward higher and lower energies until the intensity dropped below $5 \%$ of the peak height or the half-width reached 1.0 eV. Within this window, the spectral width was evaluated from the root-mean-square (RMS) deviation of the energy distribution,
\begin{equation}
\begin{aligned}
 &\sigma_{X}^\mathrm{B}=\sqrt{\frac{\int_{E_1}^{E_2}\left(E-E_{\text {cent }, X}\right)^2 w_X(E) d E}{\int_{E_1}^{E_2} w_X(E) d E}}, \quad \\&E_{\text {cent }, i}=\frac{\int_{E_1}^{E_2} E w_X(E) d E}{\int_{E_1}^{E_2} w_X(E) d E},
\end{aligned}
\end{equation}
where $w_X(E)$ is the PDOS intensity used as a weighting function.
A smaller $\sigma_{X}^\mathrm{B}$ indicates a more sharply defined resonance that is weakly hybridized with bulk or adsorbate states, whereas a larger value reflects stronger coupling and spectral broadening.

The resulting $E^\mathrm{B}_{X\text{-peak}}$ and spectral widths $\sigma^\mathrm{B}_{X}$ for Ga- and N-derived surface states were compiled across all nanocluster–substrate combinations to enable systematic comparison of their energetic alignment and degree of localization. In the present framework, Ga-derived surface states positioned closer to the conduction-band minimum (CBM), together with N-derived states closer to the valence-band maximum (VBM), yield higher Interface Scores, corresponding to a more positive Ga character and a more negative N character at the uncovered surface. The spectral width $\sigma^\mathrm{B}_{X}$ provides a quantitative measure of energetic localization. A smaller $\sigma^\mathrm{B}_{X}$ indicates a sharp, well-defined surface resonance with a narrow energy distribution, consistent with a uniform and chemically specific active site. In contrast, a large $\sigma^\mathrm{B}_{X}$ reflects broadened or multiple surface states distributed over a wide energy range, indicative of disorder or trap-like states that promote carrier recombination. Consequently, lower values of $\sigma^\mathrm{B}_{X}$ contribute positively to the Interface Score in our analysis.

\subsubsection{Region B in-plane dipole $\mu_{\|}$}
\label{sec:ap_B_descrip_dipole}

\begin{figure}
\centering
\includegraphics[scale=0.8]{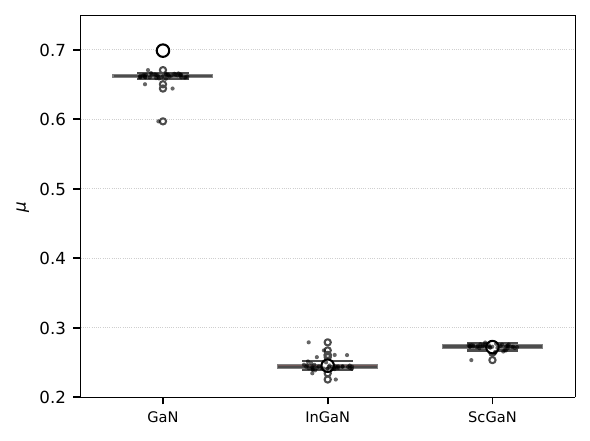}
\caption{Distribution of the in-plane dipole density $\mu^\mathrm{B}_{\|}$ of the uncovered semiconductor surface (Region B) for GaN-, InGaN-, and ScGaN-based substrates. The in-plane dipole moment per unit area is evaluated from Bader charge distributions of atoms in the top two atomic layers of Region B using Eq.~\ref{eq:mu}.  Larger $\mu^\mathrm{B}_{\|}$ indicates stronger lateral polarization, reflecting in-plane charge asymmetry induced by surface reconstruction, alloying, or intrinsic facet polarity.} 
\label{fig:inplane_dipole} 
\end{figure}

The intrinsic lateral polarization of the uncovered GaN-based surface (Region B) was evaluated from the in-plane dipole moment per unit area, $\mu_{\|}$. Using the Bader charges $q_i$ and Cartesian coordinates ($x_i, y_i$) of the surface atoms in the top two atomic layers, we first determined the geometric centroids of the surface region, $\bar{x}=\frac{1}{N} \sum_i x_i$ and $\bar{y}=\frac{1}{N} \sum_i y_i$. The magnitude of the in-plane dipole density was then obtained as
\begin{equation}
\mu_{\|}^{\text{B}} \!=\! \frac{1}{A^{\text{B}}} \! \sqrt{ \left( \sum_i q_i (x_i \!-\! \bar{x}) \right)^2 \!+\! \left( \sum_i q_i (y_i \!-\! \bar{y}) \right)^2 },
\label{eq:mu}
\end{equation}
where $A^\mathrm{B}$ is the lateral surface area of the Region $B$ patch. The sign of each component indicates the direction of the lateral electric field, while the magnitude quantifies the degree of in-plane charge asymmetry arising from surface reconstruction, alloying (In or Sc substitutions), or intrinsic polarity of the (110) facet. Larger $\mu_{\|}^\mathrm{B}$ corresponds to stronger in-plane polarization, which can facilitate lateral separation of reaction intermediates $\left(\mathrm{OH}^{-} / \mathrm{H}^{+}\right)$during the Volmer step \cite{medford2015} of the hydrogen-evolution reaction. As shown in Fig.~\ref{fig:inplane_dipole}, the in-plane dipole density varies weakly with nanocluster identity on a given substrate, indicating that facet-level polarization is primarily substrate-controlled. The enhanced in-plane dipole in GaN is attributed to its stronger intrinsic polarity and surface asymmetry on the (110) facet compared with InGaN and ScGaN, leading to stronger lateral electric fields.

\subsubsection{Region B: Charge Redistribution  \texorpdfstring{$\Delta q^{B}$}{ΔqB}}
\label{sec:ap_B_bader}
To quantify charge redistribution on the uncovered GaN surface (Region B), we
introduce the region-resolved Bader charge descriptor $\Delta q_{X}^{B}$. Specifically, $\Delta q_{X}^{B}$ is defined as the layer-resolved Bader charge
difference between the surface layer ($L_{\mathrm{surf}}$) and a deeper
reference layer ($L_{\mathrm{ref}}$) within Region B:
\begin{equation}
\Delta q^{B}_{X} = q^{B}_{X}(L_{\mathrm{surf}}) - q^{B}_{X}(L_{\mathrm{ref}}),
\end{equation}
where $q^{B}_{X}(L)$ denotes the Bader charge of atomic species $i$ (e.g., Ga or N)
in layer $L$ belonging to Region B. In this work, the surface layer is defined as layer 8, while the reference bulk-like layer is chosen as layer 4.

\subsection{Injection (Region A)}
\label{sec:ap_A_descrip}


\subsubsection{Region A Nanocluster Induced Gap States  ($N_{\mathrm{NIGS}}$)}
\label{sec:ap_A_descrip_NIGS}

Nanocluster-induced gap states (NIGS) at the nanocluster-semiconductor interface, which are analogous to metal-induced gap states (MIGS) \cite{skachkov2021, raymondt.tung2014}, were quantified from the layer-resolved PLDOS in Region A. For the first semiconductor layer beneath the interface, we integrated the PLDOS within the band-gap window $\left[E_{\mathrm{VBM}}, E_{\mathrm{CBM}}\right]$ to obtain $N^\mathrm{A}_{\text{NIGS}}$. 
\hl{Specifically, for each region and layer,} \(E_{\mathrm{VBM}}\) \hl{was defined from
the N-projected PDOS in the valence-side window} \([-6,0]\) eV \hl{as the energy at
which the cumulative integral reaches} \(95\%\) \hl{of the total N-projected PDOS
weight in that window}:
\[
\frac{
\int_{-6}^{E_{\mathrm{VBM}}}
\rho_{\mathrm{N}}(E)\,dE
}{
\int_{-6}^{0}
\rho_{\mathrm{N}}(E)\,dE
}
=
0.95 .
\]
\hl{Similarly}, \(E_{\mathrm{CBM}}\) \hl{was defined from the Ga/Sc/In-projected PDOS in
the conduction-side window} \([0,6]\) \hl{eV as the energy at which the cumulative
integral reaches} \(5\%\) \hl{of the total Ga/Sc/In-projected PDOS weight}:
\[
\frac{
\int_{0}^{E_{\mathrm{CBM}}}
\rho_{\mathrm{Ga/Sc/In}}(E)\,dE
}{
\int_{0}^{6}
\rho_{\mathrm{Ga/Sc/In}}(E)\,dE
}
=
0.05 .
\]
\hl{The NIGS analysis is described as a PDOS integration within the
gap window bounded by the reference-layer PDOS-derived} \(E_{\mathrm{VBM}}\) and
\(E_{\mathrm{CBM}}\). \hl{For example, the NIGS value in Region A, Layer 8 was
computed as}
\begin{equation}
N_{\mathrm{NIGS}}^{\mathrm{A},8}
=
\int_{E_{\mathrm{VBM}}^{\mathrm{A},6}}^
{E_{\mathrm{CBM}}^{\mathrm{A},6}}
\operatorname{PLDOS}^{\mathrm{A},8}(E)\,dE,
\label{eq:nigs}
\end{equation}
where \(E_{\mathrm{VBM}}^{\mathrm{A},6}\) and
\(E_{\mathrm{CBM}}^{\mathrm{A},6}\) \hl{are the PDOS-derived band-edge markers of
Region A, Layer 6. These reference energies are adopted because the surface layer (Layer 8) is strongly mixed with nanocluster-induced gap states, making its VBM and CBM difficult to identify. All energies in these analyses are referenced to}
\(E_{\mathrm{F}}\). 
Higher  $N^\mathrm{A}_{\text{NIGS}}$ indicates stronger pinning, worse rectification/injection. A lower $N^\mathrm{A}_{\text{NIGS}}$ denotes a “cleaner” interface with fewer trap states, allowing more efficient electron transfer to the nanocluster without carrier loss. Minimizing these states improves the rectifying contact quality and helps maintain an unpinned Fermi level for optimal band bending.

\begin{figure*}
    \centering
    \includegraphics[scale=0.8]{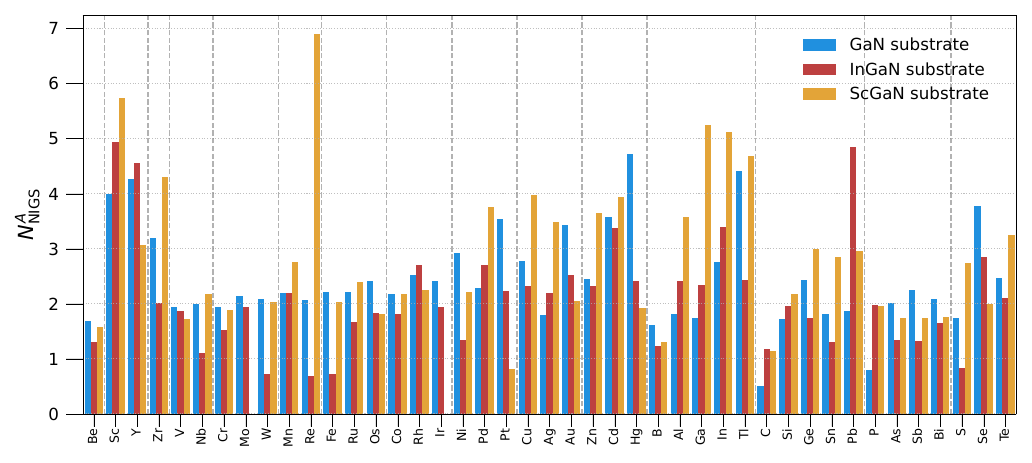}
    \caption{Nanocluster-induced gap states (NIGS) in Region A (first semiconductor layer beneath the nanocluster-semiconductor interface).
    The quantity $N^\mathrm{A}_{\text{NIGS}}$ represents the integrated projected local density of states (PLDOS) within the bandgap window $\left[E_{\mathrm{VBM}}, E_{\mathrm{CBM}}\right]$, as defined in Eq.~\ref{eq:nigs}. 
    Bars correspond to GaN (blue), InGaN (red), and ScGaN (yellow) substrates.
    A higher $N_{\text {NIGS }}^{A}$ indicates stronger metal-induced hybridization and Fermi-level pinning at the interface. Distinct substrate- and element-dependent variations highlight how interfacial NIGS evolve with alloy composition and cocatalyst identity. }
    \label{fig:nigs}
\end{figure*}

\subsubsection{Region A Band-bending: $BB_{\mathrm{CBM}}^\mathrm{A}$, $BB_{\mathrm{VBM}}^\mathrm{A}$}
\label{sec:ap_A_descrip_BB}

To quantify the interfacial electric field beneath the nanocluster, we define a Region-A band-bending descriptor based on the layer-resolved projected local density of states (PLDOS). The conduction-(valence-) band edge energy, $E_{\mathrm{CBM}}^\mathrm{A}(L)\left(E_{\mathrm{VBM}}^\mathrm{A}(L)\right)$, is identified for each layer from the Ga- (N-) derived band edges (\hl{See the mathematical definitions in  Appendix}~\ref{sec:ap_A_descrip_NIGS}). The local bending is evaluated as the energy difference between the interface layer $L_{\text{int}}$ and the reference bulk-like layer $L_{\text {ref}}$:
\begin{equation}
\begin{aligned}
&
BB_{\mathrm{CBM}}^\mathrm{A}=E_{\mathrm{CBM}}^\mathrm{A}\left(L_{\mathrm{ref}}\right)-E_{\mathrm{CBM}}^\mathrm{A}\left(L_{\mathrm{int}}\right),
\\ &
BB_{\mathrm{VBM}}^\mathrm{A}=E_{\mathrm{VBM}}^\mathrm{A}\left(L_{\mathrm{ref}}\right)-E_{\mathrm{VBM}}^\mathrm{A}\left(L_{\mathrm{int}}\right).
\end{aligned}
\end{equation}
In this work, layer 4 is taken as the bulk reference ($L_{\text {ref}}$) and layer 6 as the interface layer ($L_{\text {int}}$). These descriptors indicate the built-in electric field at the interface due to the nanocluster–semiconductor contact. Strong band bending (a large downward shift of the CBM and a large downward shift of the VBM near the interface) pushes photoexcited electrons toward the nanocluster and pushes holes into the GaN bulk. In this work, greater (and positive) $BB_{\mathrm{CBM}}^\mathrm{A}$ (and $BB_{\mathrm{VBM}}^\mathrm{A}$) generally means a stronger internal field that aids electron injection into the cluster (and holes back to the GaN bulk).

\begin{figure*}
    \centering
    \includegraphics[scale=0.7]{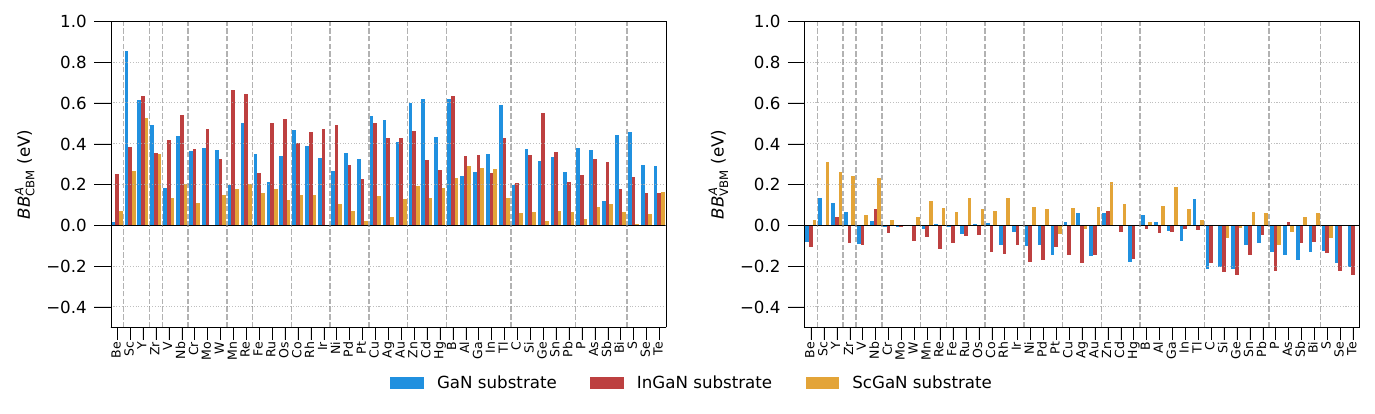}
    \caption{Element-resolved band bending at the interface region (Region A) for the conduction band minimum ($BB^\mathrm{A}_{\mathrm{CBM}}$, left) and valence band maximum ($BB^\mathrm{A}_{\mathrm{VBM}}$, right) across GaN, InGaN, and ScGaN substrates. Positive (negative) values indicate downward (upward) band bending relative to the bulk reference, reflecting substrate-dependent interfacial electrostatics and charge redistribution induced by different nanocluster compositions.} 
    \label{fig:region_BB}
\end{figure*}

\subsubsection{Region A: Charge Redistribution \texorpdfstring{$\Delta q^{A}$}{ΔqA}}
\label{sec:ap_A_Bader}
Analogous to the definition of $\Delta q_{X}^{B}$ in Appendix~\ref{sec:ap_B_bader}, we define the region-resolved Bader charge descriptor $\Delta q_{X}^{A}$ to quantify charge
redistribution in the nanocluster-covered region (Region A):
\begin{equation}
\Delta q^\mathrm{A}_{X} = q^\mathrm{A}_{X}(L_{\mathrm{surf}}) - q^\mathrm{A}_{X}(L_{\mathrm{ref}}),
\end{equation}
where $q^\mathrm{A}_{X}(L)$ denotes the Bader charge of atomic species $i$ in layer $L$
within Region A. The same surface and reference layers are used as in Region B where the surface layer is defined as layer 8 and the reference bulk-like
layer is defined as layer 4.

This descriptor captures nanocluster-induced charge accumulation or depletion
near the surface and directly reflects the efficiency of carrier injection and
electrostatic screening beneath the cocatalyst.

\subsection{Interfacial Coupling (A $\leftrightarrow$ B) }
\label{sec:ap_iface_descrip}

\subsubsection{Layer-resolved Bader charge difference between Regions A and B}
\label{sec:ap_iface_descrip_bader}

While $\Delta q_{X}^\mathrm{A}$ (Sec.~\ref{sec:ap_A_Bader}) and $\Delta q_{X}^\mathrm{B}$ (Sec.~\ref{sec:ap_B_bader}) independently
characterize charge redistribution within the nanocluster-covered and uncovered
regions, photocatalytic functionality ultimately depends on how these regions
are electrostatically coupled. To capture this lateral interaction, we introduce
interfacial coupling descriptors that quantify charge transfer and potential
offsets across the Region A-Region B junction.

The interfacial charge-transfer descriptor $\Delta q_{X}^\mathrm{AB}$ is defined as the
difference between region-resolved charge accumulations:
\begin{equation}
\Delta q^\mathrm{AB}_{X} = q^\mathrm{A}_{X}(L_{\mathrm{surf}}) - q^\mathrm{B}_{X}(L_{\mathrm{surf}}),
\end{equation}
which directly measures the degree of lateral charge imbalance bridging the two
regions on the surface. In this work, the surface layer corresponds to layer~8.

A larger Bader charge contrast signifies stronger interfacial electronic coupling.
In particular, enhanced charge transfer between the semiconductor surface and the nanocluster
indicates the formation of an effective Schottky-like contact with a built-in charge reservoir, where electrons are transferred from GaN into the nanocluster.
In the context of the hydrogen evolution reaction (HER), such charge accumulation implies that
the nanocluster is pre-loaded with electrons, facilitating subsequent proton reduction.

\begin{figure*}
\centering
\includegraphics[scale=0.8]{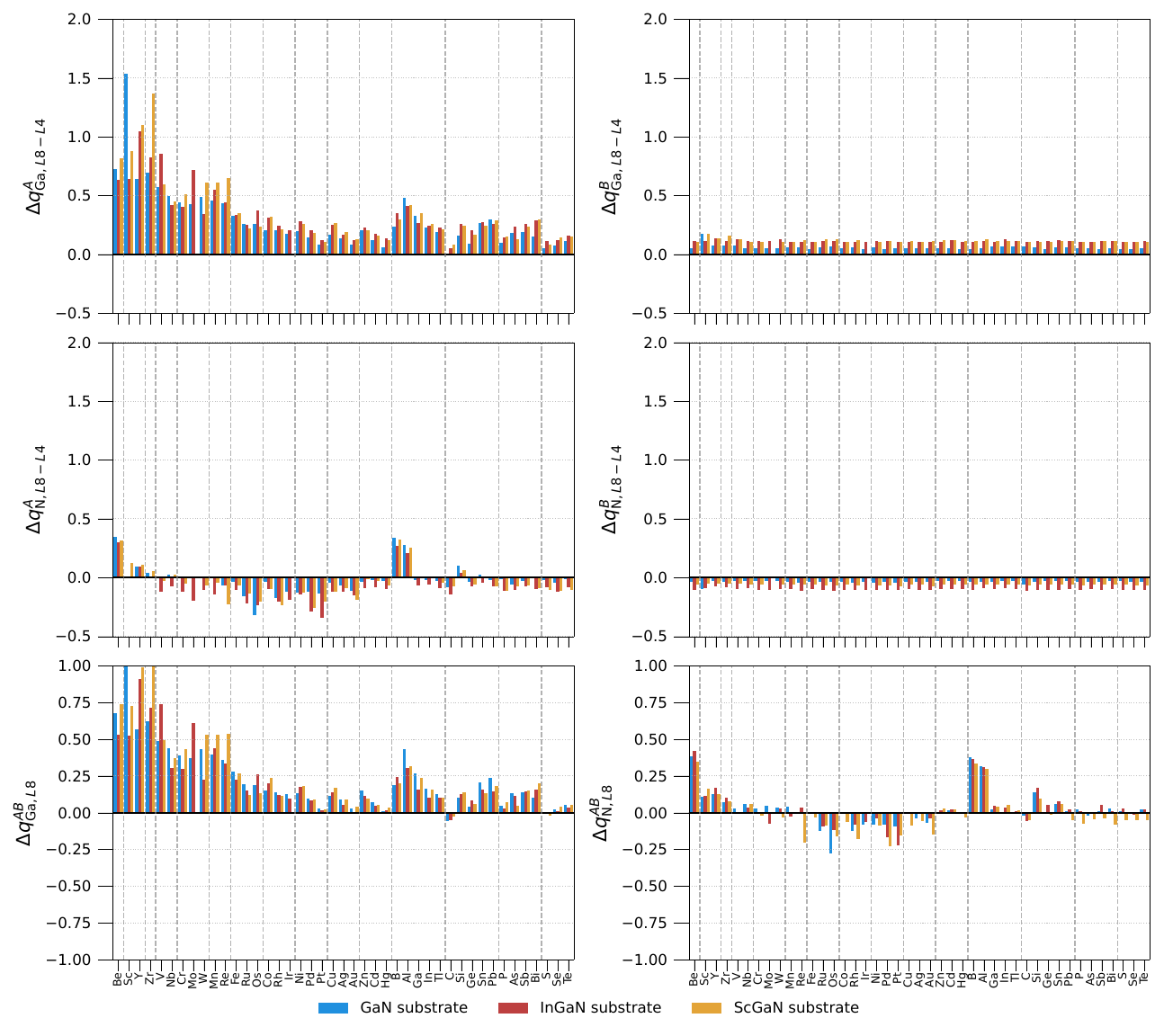}
\caption{Panels (a–d) show the layer-resolved Bader charge difference $\Delta q=q({\mathrm{L8}})-q({\mathrm{L4}})$ for Ga (a,b) and N (c,d) atoms in the nanocluster-covered Region A (a,c) and uncovered Region B (b,d). Panels (e,f) report the inter-regional surface-layer contrast at Layer 8, $\Delta q^\mathrm{AB}=q^\mathrm{A}({\mathrm{L}8})-q^\mathrm{B}({\mathrm{L}8)}$, for Ga (e) and N (f), highlighting lateral charge asymmetry induced by nanocluster adsorption. Positive (negative) values indicate electron accumulation (depletion).}
\label{fig:bader_deltas}
\end{figure*}

As shown in Fig.~\ref{fig:bader_deltas}, the nanocluster-covered Region~A exhibits a pronounced increase in Bader charge contrasts compared with the uncovered Region~B.
Within the Interface Score (IS) framework, these descriptors contribute positively to IS:
larger vertical charge contrasts (Layer~8--Layer~4) on both Ga and N atoms reflect stronger interfacial polarization and electronic coupling, while larger lateral surface-layer contrasts (Region~A--Region~B) indicate enhanced spatial charge separation and lateral carrier flow. Conversely, interfaces exhibiting weak layer-resolved polarization or negligible A-B charge contrast yield lower IS values, consistent with reduced charge-injection efficiency and increased recombination propensity.

\subsubsection{Interfacial band-edge difference between Regions A and B} 
\label{sec:ap_iface_descrip_BB_diff}

While the band-bending magnitude within Region A provides a measure of the internal electrostatic field (see Sec.~\ref{sec:ap_A_descrip_BB}), the relative alignment of the band edges between Regions A and B is directly relevant to lateral interface charge redistribution. To capture this effect, we define two separate descriptors that quantify the lateral potential offset at the interface:

\begin{equation}
\begin{aligned}
& \Delta E_{\text{CBM-edge}}^\mathrm{AB}=E_{\mathrm{CBM}}^\mathrm{A}\left(L_{\text {int}}\right)-E_{\mathrm{CBM}}^\mathrm{B}\left(L_{\text {int}}\right), \\
& \Delta E_{\text{VBM-edge}}^\mathrm{AB}=E_{\text{VBM}}^\mathrm{A}\left(L_{\text {int}}\right)-E_{\text{VBM}}^\mathrm{B}\left(L_{\text {int}}\right) .
\end{aligned}
\end{equation}

Here, $E_{\mathrm{CBM}}^\mathrm{A/B}\left(L_{\text{int }}\right)$ and $E_{\mathrm{VBM}}^\mathrm{A/B}\left(L_{\text {int }}\right)$ denote the conduction- and valence-band edge energies extracted from the Ga- and N-derived PLDOS of the interfacial layer (Layer 6 in this work) in the nanocluster-covered (A) and uncovered (B) regions.

The descriptor $\Delta E_{\text{CBM-edge}}^\mathrm{AB}$ reflects the lateral conduction-band offset, which drives photogenerated electrons from the bare semiconductor toward the nanocluster region. In contrast, $\Delta E_{\text{VBM-edge}}^\mathrm{AB}$ represents the lateral valence-band offset, which governs the migration of photoexcited holes toward the nanocluster region.

A larger magnitude of $\Delta E_{\text{edge}}^\mathrm{AB}$ suggests a pronounced in-plane potential difference between the cluster-covered and bare regions. For example, if the conduction band in Region A is significantly lower than in Region B, electrons will preferentially migrate into Region A (toward the nanocluster), while holes remain in Region B. This spatial separation of electrons vs. holes across the interface regions is crucial for preventing recombination and enabling concurrent redox reactions (electrons for HER at the cluster, holes for water oxidation at the bare surface).

\begin{figure*}
    \centering
    \includegraphics[scale=0.8]{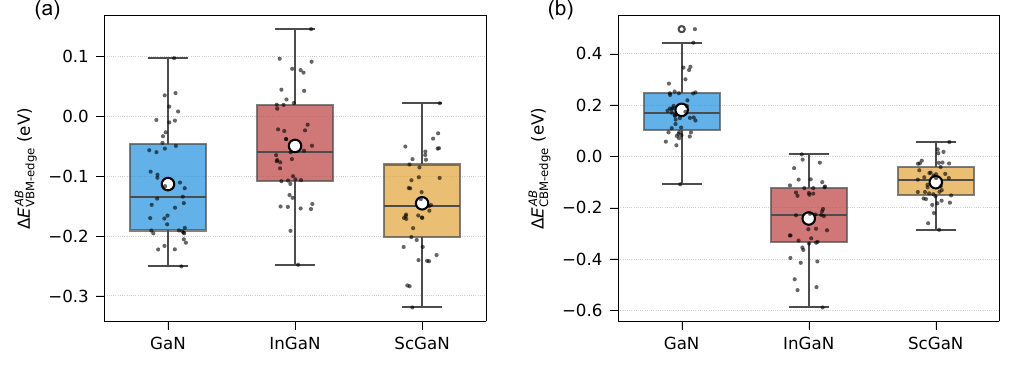}
    \caption{Distributions of the valence-band offsets \(\Delta E_{\text{VBM-edge}}^\mathrm{AB}\) (a) and  conduction-band offsets \(\Delta E_{\text{CBM-edge}}^\mathrm{AB}\) (b)  evaluated at the interfacial layer for GaN, InGaN, and ScGaN substrates. Larger offsets indicate stronger in-plane potential differences between the nanocluster-covered and uncovered regions, promoting lateral carrier separation.}
    \label{fig:Band_edge_diff}
\end{figure*}

Fig.~\ref{fig:Band_edge_diff} summarizes the lateral band-edge offsets between the nanocluster-covered Region~A and the uncovered Region~B. The conduction-band offset \(\Delta E_{\text{CBM-edge}}^\mathrm{AB}\) exhibits a pronounced substrate dependence, with larger values indicating a stronger in-plane driving force for electron migration toward the nanocluster region. In contrast, the valence-band offset \(\Delta E_{\text{VBM-edge}}^\mathrm{AB}\) is generally smaller in magnitude, suggesting a weaker but non-negligible lateral potential gradient for hole transport.

Together, these quantities provide a direct measure of the potential asymmetry between the covered and uncovered regions, offering microscopic insight into the directionality and efficiency of interfacial carrier separation during photocatalysis.

\section{Interface subscores ($I_{\mathrm{inj}}$, $I_{\text{act}}$ and $I_{\text{coupl}}$)}
\label{sec:ap_IS_score}
\setcounter{figure}{0}
\renewcommand{\thefigure}{C\arabic{figure}}
\setcounter{table}{0}
\renewcommand{\thetable}{C\arabic{table}}

Each descriptor is normalized to a unitless [0,1] scale using min-max scaling \hl{over the full dataset, including all substrate families.}
This ensures that the score reflects real physical trends, not differences in descriptor magnitudes or units. 

Charge injection from semiconductor substrate into the supported nanocluster is governed by two key electronic features in the covered region: the magnitudes of conduction- and valence-band bending beneath the cluster and the density of nanocluster-induced gap states (NIGS). We quantify this behavior through an injection score:
\begin{equation}
\begin{aligned}    
&    I_{\mathrm{inj}}=
    \alpha_1 \tilde{BB}_{\mathrm{CBM}}^\mathrm{A}
   +\alpha_2 \tilde{BB}_{\mathrm{VBM}}^\mathrm{A}
   -\alpha_3 \tilde{N}_{\mathrm{NIGS}}^\mathrm{A} 
\\&   +\alpha_4 \Delta q_\mathrm{Ga}^\mathrm{A}
   +\alpha_5 \Delta q_\mathrm{N}^\mathrm{A}
   \,,
\end{aligned}
\end{equation}
where $\tilde{BB}_{\mathrm{CBM}}^\mathrm{A}$ measures the downward bending of the conduction band, which lowers the energetic barrier for electron transfer into the nanocluster and thus contributes positively to charge injection. The term $\tilde{BB}_{\mathrm{VBM}}^\mathrm{A}$ reflects the downward bending of the valence band, which reduces the local hole concentration near the interface and suppresses electron–hole recombination. This increased electron availability also enhances catalytic activity and therefore enters with a positive coefficient. In contrast, $\tilde{N}_{\mathrm{NIGS}}^\mathrm{A}$ quantifies the density of nanocluster-induced gap states that act as recombination centers or trapping sites; these hinder carrier transport and contribute negatively to the score. 
Finally, $\Delta q_\mathrm{Ga}^\mathrm{A}$ and $\Delta q_\mathrm{N}^\mathrm{A}$ quantify surface–bulk charge redistribution in the nanocluster-covered region, with larger magnitudes indicating stronger interfacial polarization and electronic coupling; these effects enhance charge injection and are therefore assigned positive coefficients, with the sign already encoded in the charge difference itself.

Surface reactivity on the uncovered region of semiconductor substrate is shaped by the local electrostatic environment and the nature of surface-localized states. To capture these effects, we define an activation score:
\begin{equation}
\begin{aligned}
    I_{\text{act}}
    &=\beta_1\tilde{E}_{\text{Ga-peak}}
    +\beta_2\tilde{E}_{\text{N-peak}}
    -\beta_3\tilde{\sigma}_{\text{Ga-peak}}
 \\ &   -\beta_4\tilde{\sigma}_{\text{N-peak}} 
    +\beta_5\tilde{\mu}_{\|} 
   +\beta_6 \Delta q_\mathrm{Ga}^\mathrm{B}
    +\beta_7 \Delta q_\mathrm{N}^\mathrm{B}
\end{aligned}
\end{equation}
where $\tilde{E}_{\text{Ga-peak}}$ and $\tilde{E}_{\text{N-peak}}$ denotes the normalized energy of the dominant Ga- and N- derived surface state peaks. Higher peak energies correspond to states positioned closer to the Fermi level, which enhances the ability of the surface to accept electrons and thereby facilitates adsorbate activation.   \hl{These surface states in the uncovered region may participate in adsorbate interactions and water activation. However, the present activation score is based on static ground-state electronic descriptors. Quantitatively distinguishing beneficial adsorbate-activation states from detrimental recombination centers would require explicit solvent passivation and excited-state carrier-capture calculations, which remain important future extensions of this framework.} The quantity $\tilde{\sigma}_{\text{Ga-peak}}$ and $\tilde{\sigma}_{\text{N-peak}}$ represents the normalized spectral width of the peaks; broader widths indicate more dispersed electronic density, diminishing the localization required for strong binding interactions and thus enter with a negative sign. The term $\tilde{\mu}_{\|}$ is the normalized in-plane dipole moment, capturing lateral electrostatic gradients at the surface. These gradients polarize interfacial water molecules and promote proton activation, contributing positively to the overall surface reactivity. Finally, $\Delta q_\mathrm{Ga}^\mathrm{B}$ and $\Delta q_\mathrm{N}^\mathrm{B}$ quantify surface–bulk charge redistribution in Region~B; larger magnitudes indicate stronger surface polarization and enhanced electrostatic interaction with adsorbates, and therefore contribute positively to the activation score, with the sign already encoded in the charge difference itself.

The covered and uncovered regions together form a lateral heterojunction, whose behavior is determined by how charge redistributes across the interface and how the corresponding band edges align. To capture these effects, we define a coupling score
\begin{equation}
    I_{\text{coupl}}\!=\!
    \gamma_1 \Delta E_{\mathrm{CBM}}^\mathrm{AB}
   \!+\!\gamma_2 \Delta E_{\mathrm{VBM}}^\mathrm{AB}
   \!+\!\gamma_3 \Delta q^{\mathrm{AB}}_\mathrm{Ga}
   \!+\!\gamma_4 \Delta q^{\mathrm{AB}}_\mathrm{N}
\end{equation}
where the terms $E_\mathrm{CBM}^\mathrm{AB}$ and $\Delta E_\mathrm{VBM}^\mathrm{AB}$ denote the normalized CBM and VBM band-edge offsets across the lateral junction. Larger charge transfer enhances lateral carrier flow and strengthens spatial charge separation, while favorable band offsets facilitate directional carrier migration between the two regions. Both effects contribute positively to the overall surface reactivity. $\Delta q^{\mathrm{AB}}_{\mathrm{Ga}}$ and $\Delta q^{\mathrm{AB}}_{\mathrm{N}}$ represent the normalized interfacial charge transfer on Ga and N atoms between Regions A and B, respectively.

\section{Interface Score calibration and representative outcomes}
\label{sec:ap_IS_score_cali_example}
\setcounter{figure}{0}
\renewcommand{\thefigure}{D\arabic{figure}}
\setcounter{table}{0}
\renewcommand{\thetable}{D\arabic{table}}

The Interface Score (IS) weighting parameters were calibrated using a randomized search over the non-negative weight space of all normalized interface descriptors listed in Table~\ref{tab:iface} and detailed in Appendix~\ref{sec:ap_interface_descriptors}. 
Here, the randomized search denotes repeated sampling of non‑negative weight vectors from a Dirichlet distribution, followed by scaling so that $\sum_i w_i = 2.0$, with no additional structural constraints imposed. All descriptors were normalized to the interval $[0,1]$, with fixed sign conventions indicating whether a given descriptor promotes or penalizes interfacial efficiency.
The optimization objective followed a weakly constrained criterion, requiring that at least one Rh-containing interface attain the highest rank among all candidate systems, while allowing both substrate identity and internal descriptor weights to vary freely. The calibration was repeated over 20 independent random seeds, each sampling up to $8\times10^5$ candidate parameter sets. In all runs, a Rh-based interface achieved rank~1, corresponding to a 100\% success rate (Table~\ref{tab:robustness}), indicating that the resulting ranking is robust with respect to random sampling (random seeds). One representative optimized parameterization (corresponding to seed~1) is reported below to illustrate the relative descriptor contributions (Table~\ref{tab:weights}) and the resulting top 10 interface ranking (Table~\ref{tab:top10}). We emphasize that this parameterization is not unique; multiple weighting combinations yield equivalent ranking performance, and the reported values are intended to illustrate representative trends rather than to assert definitive mechanistic dominance.

\begin{table}
\caption{\label{tab:robustness}Robustness statistics of the Interface Score calibration across 20 independent random seeds.}
\begin{ruledtabular}
\begin{tabular}{lc}
\noalign{\smallskip}
Quantity & Value \\
\hline
\noalign{\smallskip}
Number of random seeds & 20 \\
Success rate ($Rh$ rank = 1) & 100\% \\
Best $Rh$ rank & 1 \\
Mean best-$Rh$ rank & 1.00 \\
Worst best-$Rh$ rank & 1 \\
\noalign{\smallskip}
\end{tabular}
\end{ruledtabular}
\end{table}

\begin{table*}
\caption{\label{tab:weights}Representative optimized Interface Score weights (seed~1), shown with hierarchical decomposition into injection, activation, and coupling contributions.}
\begin{ruledtabular}
\begin{tabular}{lllc}
\noalign{\smallskip}
Component & Descriptor & Physical meaning & Weight \\
\hline
\noalign{\smallskip}
\multicolumn{4}{l}{\textit{Injection} ($I_{\mathrm{inj}}$)} \\
& $E_{\mathrm{VBM}}^{\mathrm{A}}$ & Region A VBM band bending & 0.6149 \\
& NIGS density & Nanocluster induced gap states & 0.0100 \\
& $E_{\mathrm{CBM}}^{\mathrm{A}}$ & Region A CBM band bending & 0.0005 \\
& $\Delta q_{\mathrm{Ga}}^{\mathrm{A}}$ & Bader charge on Ga (Region A) & 0.0493 \\
& $\Delta q_{\mathrm{N}}^{\mathrm{A}}$ & Bader charge on N (Region A) & 0.0409 \\
\noalign{\medskip}
\multicolumn{4}{l}{\textit{Activation} ($I_{\mathrm{act}}$)} \\
& $E_{\text{peak}}^{\mathrm{N}}$ & N-derived surface state energy & 0.3191 \\
& $E_{\text{peak}}^{\mathrm{Ga}}$ & Ga-derived surface state energy & 0.2195 \\
& $\mu$ & In-plane dipole magnitude & 0.0988 \\
& $\sigma_{\mathrm{Ga}}$ & Peak broadening penalty & 0.0600 \\
& $\sigma_{\mathrm{N}}$ & Peak broadening penalty & 0.0154 \\
& $\Delta q_{\mathrm{Ga}}^{\mathrm{B}}$ & Bader charge on Ga (Region B) & 0.1855 \\
& $\Delta q_{\mathrm{N}}^{\mathrm{B}}$ & Bader charge on N (Region B) & 0.0959 \\
\noalign{\medskip}
\multicolumn{4}{l}{\textit{Coupling} ($I_{\mathrm{coupl}}$)} \\
& $\Delta E_{\mathrm{CBM}}^{\mathrm{AB}}$ & CBM offset across interface & 0.2229 \\
& $\Delta E_{\mathrm{VBM}}^{\mathrm{AB}}$ & VBM offset across interface & 0.0342 \\
& $\Delta q_{\mathrm{N}}^{\mathrm{AB}}$ & Interfacial N charge transfer & 0.0288 \\
& $\Delta q_{\mathrm{Ga}}^{\mathrm{AB}}$ & Interfacial Ga charge transfer & 0.0042 \\
\hline
\noalign{\smallskip}
\multicolumn{4}{l}{Normalized outer weights: $w_A=0.358$, $w_B=0.497$, $w_{AB}=0.145$} \\
\end{tabular}
\end{ruledtabular}
\end{table*}

\begin{table}[h]
\caption{\label{tab:top10}Top ten interfaces ranked by the representative optimized Interface Score.}
\begin{ruledtabular}
\begin{tabular}{cllc}
\noalign{\smallskip}
Rank & Element & Substrate & IS \\
\hline
\noalign{\smallskip}
1  & Rh & ScGaN & 1.067 \\
2  & Se & ScGaN & 1.032 \\
3  & Au & ScGaN & 1.029 \\
4  & Te & ScGaN & 0.994 \\
5  & C  & ScGaN & 0.992 \\
6  & Al & ScGaN & 0.990 \\
7  & S  & ScGaN & 0.968 \\
8  & Pd & ScGaN & 0.963 \\
9  & Zr & ScGaN & 0.956 \\
10 & Sc & GaN   & 0.955 \\
\noalign{\smallskip}
\end{tabular}
\end{ruledtabular}
\end{table}

\section{Machine-learning validation controls}
\label{app:ml_validation_controls}

To assess robustness beyond the fixed 80/20 holdout split, we performed additional validation tests including 50 repeated random 80/20 train/test splits, leave-one-element-out validation, leave-one-substrate-out validation, and random-label controls. Random-label controls were performed by shuffling the H-adsorption-energy labels before training, thereby destroying the true descriptor--property relationship.

For the best global-descriptor model, the repeated-split validation gives $R^2 = 0.779 \pm 0.088$ and RMSE $= 0.308 \pm 0.062$ eV. Leave-one-element-out validation gives $R^2 = 0.692$ and RMSE $= 0.402$ eV, while leave-one-substrate-out validation gives $R^2 = 0.787$ and RMSE $= 0.335$ eV. In contrast, the random-label control gives negative mean performance, $R^2 = -0.325 \pm 0.335$, confirming that comparable accuracy is not obtained when the target-property relationship is destroyed.

The same validation protocol was also applied to the best interface-descriptor model. The interface-only model gives lower predictive accuracy than the global-descriptor model, but remains above the random-label baseline. This supports our use of global descriptors for quantitative H-adsorption-energy regression, while treating interface descriptors primarily as physically interpretable local descriptors for the separate Interface Score analysis.

\begin{table*}[h!]
\centering
\caption{
Robustness validation of the H-adsorption-energy regression models. 
Repeated 80/20 validation reports mean $\pm$ standard deviation over 50 random splits. 
Random-label controls were performed by shuffling the H-adsorption-energy labels before training. 
Leave-one-element-out and leave-one-substrate-out validations report out-of-fold predictions over all samples.
}
\label{tab:ml_validation_controls}
\begin{tabular}{llcccc}
\hline
Descriptor set & Validation protocol & $R^2$ & MAE (eV) & RMSE (eV) & Splits \\
\hline
Global & Repeated 80/20 split 
& $0.779 \pm 0.088$ & $0.236 \pm 0.044$ & $0.308 \pm 0.062$ & 50 \\
Global & Leave-one-element-out 
& $0.692$ & $0.292$ & $0.402$ & 43 \\
Global & Leave-one-substrate-out 
& $0.787$ & $0.254$ & $0.335$ & 3 \\
Global & Random-label 80/20 split 
& $-0.325 \pm 0.335$ & $0.642 \pm 0.090$ & $0.804 \pm 0.106$ & 50 \\
\hline
Interface & Repeated 80/20 split 
& $0.354 \pm 0.184$ & $0.424 \pm 0.063$ & $0.534 \pm 0.088$ & 50 \\
Interface & Leave-one-element-out 
& $0.393$ & $0.433$ & $0.564$ & 43 \\
Interface & Leave-one-substrate-out 
& $0.290$ & $0.480$ & $0.610$ & 3 \\
Interface & Random-label 80/20 split 
& $-0.200 \pm 0.179$ & $0.615 \pm 0.084$ & $0.773 \pm 0.094$ & 50 \\
\hline
\end{tabular}
\end{table*}

\section{Hyperparameter Search Space}
\label{app:hyperparams}
\setcounter{figure}{0}
\renewcommand{\thefigure}{E\arabic{figure}}
\setcounter{table}{0}
\renewcommand{\thetable}{E\arabic{table}}

This appendix summarizes the randomized search space used for each model (Table~\ref{tab:searchCV}.). The
same search space was applied to both the global and interface descriptor sets. Best hyperparameters for all models and descriptor sets are shown in Table~\ref{tab:bestCV}.

\begin{table*}
\caption{\label{tab:searchCV}RandomizedSearchCV hyperparameter space for each model.}
\begin{ruledtabular}
\begin{tabular}{ll}
\noalign{\smallskip}
\textbf{Model} & \textbf{RandomizedSearchCV parameter space} \\
\hline
\noalign{\medskip}
RF & 
\parbox[c]{12cm}{\raggedright n\_estimators $\in \{200, 500, 800\}$; max\_depth $\in \{\text{None}, 5, 10, 20\}$; max\_features $\in \{\text{sqrt}, \text{None}, 0.5\}$; min\_samples\_leaf $\in \{1, 2, 4\}$.} \\
\noalign{\medskip}
ET & 
\parbox[c]{12cm}{\raggedright n\_estimators $\in \{200, 500, 800\}$; max\_depth $\in \{\text{None}, 5, 10, 20\}$; max\_features $\in \{\text{sqrt}, \text{None}, 0.5\}$; min\_samples\_leaf $\in \{1, 2, 4\}$.} \\
\noalign{\medskip}
GBR & 
\parbox[c]{12cm}{\raggedright n\_estimators $\in \{200, 500\}$; learning\_rate $\in \{0.05, 0.1, 0.2\}$; max\_depth $\in \{2, 3, 4\}$; subsample $\in \{0.7, 1.0\}$; min\_samples\_leaf $\in \{1, 3, 5\}$.} \\
\noalign{\medskip}
HGB & 
\parbox[c]{12cm}{\raggedright max\_iter $\in \{600, 1000\}$; learning\_rate $\in \{0.02, 0.05\}$; max\_depth $\in \{3, 5, \text{None}\}$; max\_leaf\_nodes $\in \{31, 63, 127\}$; min\_samples\_leaf $\in \{20, 30, 50\}$; l2\_regularization $\in \{10^{-3}, 10^{-2}, 10^{-1}\}$; early\_stopping $\in \{\text{True}\}$; validation\_fraction $\in \{0.2\}$; n\_iter\_no\_change $\in \{10, 20\}$; max\_bins $\in \{64, 128, 255\}$.} \\
\noalign{\medskip}
SVR & 
\parbox[c]{12cm}{\raggedright $C \in \{10^{-2}, 10^{-1}, 1, 10, 10^{2}, 10^{3}\}$; $gamma \in \{\text{scale}, 10^{-4}, 10^{-3}, 10^{-2}, 10^{-1}, 1.0\}$; $epsilon \in \{0.01, 0.05, 0.1, 0.2\}$; $kernel \in \{\text{rbf}\}$; optional PCA with n\_components $\in \{0.8, 0.9, 0.95, \text{None}\}$.} \\
\noalign{\medskip}
LASSO & 
\parbox[c]{12cm}{\raggedright $alpha \in \{10^{-4}, 3\times 10^{-4}, 10^{-3}, 3\times 10^{-3}, 10^{-2}, 3\times 10^{-2}, 10^{-1}, 0.3, 1, 3\}$.} \\
\noalign{\smallskip}
\end{tabular}
\end{ruledtabular}
\end{table*}
\newpage

\begin{table*}
\caption{\label{tab:bestCV}Best hyperparameters for all models and descriptor sets.}
\begin{ruledtabular}
\begin{tabular}{lll}
\noalign{\smallskip}
\textbf{Descriptor set} & \textbf{Model} & \textbf{Best hyperparameters} \\
\hline
\noalign{\medskip}
Global & LASSO & \parbox[c]{10cm}{\raggedright alpha = 0.03.} \\
\noalign{\smallskip}
Interface & LASSO & \parbox[c]{10cm}{\raggedright alpha = 0.1.} \\
\noalign{\medskip}
Global & RF & \parbox[c]{10cm}{\raggedright max\_depth = 20; max\_features = 0.5; min\_samples\_leaf = 1; n\_estimators = 500.} \\
\noalign{\smallskip}
Interface & RF & \parbox[c]{10cm}{\raggedright max\_depth = 10; max\_features = 0.5; min\_samples\_leaf = 2; n\_estimators = 800.} \\
\noalign{\medskip}
Global & ET & \parbox[c]{10cm}{\raggedright max\_depth = None; max\_features = None; min\_samples\_leaf = 1; n\_estimators = 500.} \\
\noalign{\smallskip}
Interface & ET & \parbox[c]{10cm}{\raggedright max\_depth = 20; max\_features = None; min\_samples\_leaf = 2; n\_estimators = 800.} \\
\noalign{\medskip}
Global & GBR & \parbox[c]{10cm}{\raggedright learning\_rate = 0.1; max\_depth = 4; min\_samples\_leaf = 1; n\_estimators = 500; subsample = 0.7.} \\
\noalign{\smallskip}
Interface & GBR & \parbox[c]{10cm}{\raggedright learning\_rate = 0.1; max\_depth = 2; min\_samples\_leaf = 1; n\_estimators = 200; subsample = 0.7.} \\
\noalign{\medskip}
Global & HGB & \parbox[c]{10cm}{\raggedright early\_stopping = True; l2\_regularization = 0.01; learning\_rate = 0.05; max\_bins = 64; max\_depth = 5; max\_iter = 1000; max\_leaf\_nodes = 63; min\_samples\_leaf = 20; n\_iter\_no\_change = 20; validation\_fraction = 0.2.} \\
\noalign{\smallskip}
Interface & HGB & \parbox[c]{10cm}{\raggedright early\_stopping = True; l2\_regularization = 0.1; learning\_rate = 0.02; max\_bins = 128; max\_depth = 3; max\_iter = 600; max\_leaf\_nodes = 127; min\_samples\_leaf = 20; n\_iter\_no\_change = 20; validation\_fraction = 0.2.} \\
\noalign{\medskip}
Global & SVR & \parbox[c]{10cm}{\raggedright $C = 100$; epsilon = 0.2; gamma = 0.01; kernel = rbf; pca\_\_n\_components = 0.95.} \\
\noalign{\smallskip}
Interface & SVR & \parbox[c]{10cm}{\raggedright $C = 1$; epsilon = 0.05; gamma = 0.01; kernel = rbf; pca\_\_n\_components = 0.8.} \\
\noalign{\smallskip}
\end{tabular}
\end{ruledtabular}
\end{table*}

\section{Mathematical Definition of Region and Layer Masks Used for LDOS Analysis}

All region definitions were made in Cartesian coordinates obtained from the
POSCAR structure. Let atom \(i\) have Cartesian position
\(\mathbf{r}_i=(x_i,y_i,z_i)\). Let \(C\) denote the set of nanocluster atoms.
Region A was defined as the in-plane rectangular mask enclosing the
nanocluster, expanded by \(1.0~\text{\AA}\) in both lateral directions:
\[
\mathrm{A}
=
\left\{
i :
x_{\mathrm{A}}^{-} \le x_i \le x_{\mathrm{A}}^{+},
\quad
y_{\mathrm{A}}^{-} \le y_i \le y_{\mathrm{A}}^{+}
\right\},
\]
where
\[
x_{\mathrm{A}}^{-}=\min_{j\in C} x_j - 1.0~\text{\AA}, \qquad
x_{\mathrm{A}}^{+}=\max_{j\in C} x_j + 1.0~\text{\AA},
\]
\[
y_{\mathrm{A}}^{-}=\min_{j\in C} y_j - 1.0~\text{\AA}, \qquad
y_{\mathrm{A}}^{+}=\max_{j\in C} y_j + 1.0~\text{\AA}.
\]

Region B was defined as the exposed surface region outside Region A but
inside the full in-plane surface mask. Let \(S\) be the rectangular mask
spanning all atoms in the structure, expanded by \(2.0~\text{\AA}\):
\[
S
=
\left\{
i :
x_{\mathrm{S}}^{-} \le x_i \le x_{\mathrm{S}}^{+},
\quad
y_{\mathrm{S}}^{-} \le y_i \le y_{\mathrm{S}}^{+}
\right\},
\]
with
\[
x_{\mathrm{S}}^{-}=\min_i x_i - 2.0~\text{\AA}, \qquad
x_{\mathrm{S}}^{+}=\max_i x_i + 2.0~\text{\AA},
\]
\[
y_{\mathrm{S}}^{-}=\min_i y_i - 2.0~\text{\AA}, \qquad
y_{\mathrm{S}}^{+}=\max_i y_i + 2.0~\text{\AA}.
\]
Region B was then defined as
\[
\mathrm{B}=S\setminus \mathrm{A}.
\]
Thus, the lateral boundary between Regions A and B is the rectangular
boundary of Region A, given by
\[
x=x_{\mathrm{A}}^{-},\quad x=x_{\mathrm{A}}^{+},\quad
y=y_{\mathrm{A}}^{-},\quad y=y_{\mathrm{A}}^{+}.
\]
Atoms lying exactly on this boundary were assigned to Region A and excluded
from Region B.

Layer indices were assigned after applying the corresponding regional mask.
For each region, atoms were sorted by their Cartesian \(z\) coordinate. A new
layer was started whenever the spacing between two adjacent sorted
\(z\)-coordinates exceeded the tolerance
\(\Delta z_{\mathrm{tol}}=0.5~\text{\AA}\):
\[
z_{k+1}-z_k > \Delta z_{\mathrm{tol}}.
\]
The resulting layers were numbered sequentially from the lowest-\(z\) layer
to the highest-\(z\) layer. Therefore, layer index \(1\) corresponds to the
bottom-most atomic layer within the selected region, and larger layer indices
correspond to atoms at increasing height along the surface normal.

\newpage

\bibliography{Paper5_GaN_ML}

@article{alqahtani2016,
  title = {Atomically Resolved Structure of Ligand-Protected {{Au9}} Clusters on {{TiO2}} Nanosheets Using Aberration-Corrected {{STEM}}},
  author = {Al Qahtani, Hassan S. and Kimoto, Koji and Bennett, Trystan and Alvino, Jason F. and Andersson, Gunther G. and Metha, Gregory F. and Golovko, Vladimir B. and Sasaki, Takayoshi and Nakayama, Tomonobu},
  year = 2016,
  month = mar,
  journal = {The Journal of Chemical Physics},
  volume = {144},
  number = {11},
  pages = {114703},
  issn = {0021-9606, 1089-7690},
  doi = {10.1063/1.4943203},
  urldate = {2024-04-12},
  langid = {english}
}

@article{batatia2022,
  title = {{{MACE}}: {{Higher Order Equivariant Message Passing Neural Networks}} for {{Fast}} and {{Accurate Force Fields}}},
  author = {Batatia, Ilyes and Kovacs, David P and Simm, Gregor and Ortner, Christoph and Csanyi, Gabor},
  editor = {Koyejo, S. and Mohamed, S. and Agarwal, A. and Belgrave, D. and Cho, K. and Oh, A.},
  year = 2022,
  month = dec,
  journal = {Advances in Neural Information Processing Systems},
  volume = {35},
  pages = {11423--11436},
  doi = {10.48550/arXiv.2206.07697}
}

@article{Blochl1994,
  title = {Projector Augmented-Wave Method},
  author = {Bl{\"o}chl, P. E.},
  year = 1994,
  month = dec,
  journal = {Physical Review B},
  volume = {50},
  number = {24},
  eprint = {1408.4701v2},
  pages = {17953--17979},
  publisher = {American Physical Society},
  issn = {01631829},
  doi = {10.1103/PhysRevB.50.17953},
  urldate = {2017-12-14},
  archiveprefix = {arXiv},
  isbn = {0163-1829 (Print)\textbackslash n0163-1829 (Linking)},
  pmid = {9976227}
}

@article{che2013,
  title = {Nobel {{Prize}} in Chemistry 1912 to {{Sabatier}}: {{Organic}} Chemistry or Catalysis?},
  shorttitle = {Nobel {{Prize}} in Chemistry 1912 to {{Sabatier}}},
  author = {Che, Michel},
  year = 2013,
  month = dec,
  journal = {Catalysis Today},
  series = {Catalysis: {{From}} the Active Sites to the Processes},
  volume = {218--219},
  pages = {162--171},
  issn = {0920-5861},
  doi = {10.1016/j.cattod.2013.07.006},
  urldate = {2026-01-17}
}

@article{chen2024,
  title = {Unusual {{Sabatier}} Principle on High Entropy Alloy Catalysts for Hydrogen Evolution Reactions},
  author = {Chen, Zhi Wen and Li, Jian and Ou, Pengfei and Huang, Jianan Erick and Wen, Zi and Chen, LiXin and Yao, Xue and Cai, GuangMing and Yang, Chun Cheng and Singh, Chandra Veer and Jiang, Qing},
  year = 2024,
  month = jan,
  journal = {Nat Commun},
  volume = {15},
  number = {1},
  pages = {359},
  issn = {2041-1723},
  doi = {10.1038/s41467-023-44261-4},
  urldate = {2024-11-25},
  langid = {english}
}

@article{cohen2008,
  title = {Insights into {{Current Limitations}} of {{Density Functional Theory}}},
  author = {Cohen, Aron J. and {Mori-S{\'a}nchez}, Paula and Yang, Weitao},
  year = 2008,
  month = aug,
  journal = {Science},
  volume = {321},
  number = {5890},
  pages = {792--794},
  publisher = {American Association for the Advancement of Science},
  doi = {10.1126/science.1158722},
  urldate = {2025-12-18}
}

@article{davis2018,
  title = {Net-Zero Emissions Energy Systems},
  author = {Davis, Steven J. and Lewis, Nathan S. and Shaner, Matthew and Aggarwal, Sonia and Arent, Doug and Azevedo, In{\^e}s L. and Benson, Sally M. and Bradley, Thomas and Brouwer, Jack and Chiang, Yet-Ming and Clack, Christopher T. M. and Cohen, Armond and Doig, Stephen and Edmonds, Jae and Fennell, Paul and Field, Christopher B. and Hannegan, Bryan and Hodge, Bri-Mathias and Hoffert, Martin I. and Ingersoll, Eric and Jaramillo, Paulina and Lackner, Klaus S. and Mach, Katharine J. and Mastrandrea, Michael and Ogden, Joan and Peterson, Per F. and Sanchez, Daniel L. and Sperling, Daniel and Stagner, Joseph and Trancik, Jessika E. and Yang, Chi-Jen and Caldeira, Ken},
  year = 2018,
  month = jun,
  journal = {Science},
  volume = {360},
  number = {6396},
  pages = {eaas9793},
  publisher = {American Association for the Advancement of Science},
  doi = {10.1126/science.aas9793},
  urldate = {2025-12-17}
}

@article{doan2023,
  title = {Computational {{Screening}} of {{Supported Metal Oxide Nanoclusters}} for {{Methane Activation}}: {{Insights}} into {{Homolytic}} versus {{Heterolytic C}}--{{H Bond Dissociation}}},
  shorttitle = {Computational {{Screening}} of {{Supported Metal Oxide Nanoclusters}} for {{Methane Activation}}},
  author = {Doan, Hieu A. and Wang, Xijun and Snurr, Randall Q.},
  year = 2023,
  month = jun,
  journal = {J. Phys. Chem. Lett.},
  volume = {14},
  number = {21},
  pages = {5018--5024},
  issn = {1948-7185, 1948-7185},
  doi = {10.1021/acs.jpclett.3c00863},
  urldate = {2025-09-26},
  copyright = {https://doi.org/10.15223/policy-029},
  langid = {english}
}

@article{fathabadi2025,
  title = {Scandium-{{III-nitrides}}: {{A New Material Platform}} for {{Semiconductor Photocatalysts}} with {{High Reducing Power}}},
  shorttitle = {Scandium-{{III-nitrides}}},
  author = {Fathabadi, Milad and Vafadar, Mohammad Fazel and Ni, Siting and Zhao, Ying and Song, Jun and Li, Chao-Jun and Zhao, Songrui},
  year = 2025,
  month = jan,
  journal = {Nano Lett.},
  volume = {25},
  number = {2},
  pages = {786--792},
  issn = {1530-6984, 1530-6992},
  doi = {10.1021/acs.nanolett.4c05065},
  urldate = {2025-12-18},
  copyright = {https://doi.org/10.15223/policy-029},
  langid = {english}
}

@article{gehringer2023,
  title = {Models of Configurationally-Complex Alloys Made Simple},
  author = {Gehringer, Dominik and Fri{\'a}k, Martin and Holec, David},
  year = 2023,
  month = may,
  journal = {Computer Physics Communications},
  volume = {286},
  pages = {108664},
  issn = {00104655},
  doi = {10.1016/j.cpc.2023.108664},
  urldate = {2025-01-17},
  langid = {english}
}

@article{Grimme2010,
  title = {A Consistent and Accurate Ab Initio Parametrization of Density Functional Dispersion Correction ({{DFT-D}}) for the 94 Elements {{H-Pu}}},
  author = {Grimme, Stefan and Antony, Jens and Ehrlich, Stephan and Krieg, Helge},
  year = 2010,
  month = apr,
  journal = {The Journal of Chemical Physics},
  volume = {132},
  number = {15},
  pages = {154104},
  issn = {0021-9606},
  doi = {10.1063/1.3382344},
  pmid = {20423165}
}

@article{guan2023,
  title = {Hydrogen Society: From Present to Future},
  shorttitle = {Hydrogen Society},
  author = {Guan, Daqin and Wang, Bowen and Zhang, Jiguang and Shi, Rui and Jiao, Kui and Li, Lincai and Wang, Yang and Xie, Biao and Zhang, Qingwen and Yu, Jie and Zhu, Yunfeng and Shao, Zongping and Ni, Meng},
  year = 2023,
  journal = {Energy \& Environmental Science},
  volume = {16},
  number = {11},
  pages = {4926--4943},
  publisher = {Royal Society of Chemistry},
  doi = {10.1039/D3EE02695G},
  urldate = {2025-12-17},
  langid = {english}
}

@article{gusarov2024,
  title = {Advances in {{Computational Methods}} for {{Modeling Photocatalytic Reactions}}: {{A Review}} of {{Recent Developments}}},
  shorttitle = {Advances in {{Computational Methods}} for {{Modeling Photocatalytic Reactions}}},
  author = {Gusarov, Sergey},
  year = 2024,
  month = apr,
  journal = {Materials (Basel)},
  volume = {17},
  number = {9},
  pages = {2119},
  issn = {1996-1944},
  doi = {10.3390/ma17092119},
  urldate = {2025-12-18},
  pmcid = {PMC11085804},
  pmid = {38730926}
}

@article{hammer2000,
  title = {Theoretical Surface Science and Catalysis---Calculations and Concepts},
  author = {Hammer, B. and N{\o}rskov, J.K.},
  year = 2000,
  journal = {Advances in Catalysis},
  volume = {45},
  pages = {71--129},
  publisher = {Elsevier},
  doi = {10.1016/S0360-0564(02)45013-4},
  urldate = {2024-11-06},
  langid = {english}
}

@article{hisatomi2019,
  title = {Reaction Systems for Solar Hydrogen Production via Water Splitting with Particulate Semiconductor Photocatalysts},
  author = {Hisatomi, Takashi and Domen, Kazunari},
  year = 2019,
  month = may,
  journal = {Nat Catal},
  volume = {2},
  number = {5},
  pages = {387--399},
  publisher = {Nature Publishing Group},
  issn = {2520-1158},
  doi = {10.1038/s41929-019-0242-6},
  urldate = {2025-12-17},
  copyright = {2019 Springer Nature Limited},
  langid = {english}
}

@article{Jain2013,
  title = {Commentary: {{The}} Materials Project: {{A}} Materials Genome Approach to Accelerating Materials Innovation},
  author = {Jain, Anubhav and Ong, Shyue Ping and Hautier, Geoffroy and Chen, Wei and Richards, William Davidson and Dacek, Stephen and Cholia, Shreyas and Gunter, Dan and Skinner, David and Ceder, Gerbrand and Persson, Kristin A.},
  year = 2013,
  month = jul,
  journal = {APL Materials},
  volume = {1},
  number = {1},
  pages = {011002},
  issn = {2166532X},
  doi = {10.1063/1.4812323},
  urldate = {2017-12-14},
  isbn = {2166532X}
}

@article{johnson2025,
  title = {Realistic Roles for Hydrogen in the Future Energy Transition},
  author = {Johnson, Nathan and Liebreich, Michael and Kammen, Daniel M. and Ekins, Paul and McKenna, Russell and Staffell, Iain},
  year = 2025,
  month = may,
  journal = {Nat. Rev. Clean Technol.},
  volume = {1},
  number = {5},
  pages = {351--371},
  publisher = {Nature Publishing Group},
  issn = {3005-0685},
  doi = {10.1038/s44359-025-00050-4},
  urldate = {2025-12-17},
  copyright = {2025 Springer Nature Limited},
  langid = {english}
}

@article{kamat2012,
  title = {Manipulation of {{Charge Transfer Across Semiconductor Interface}}. {{A Criterion That Cannot Be Ignored}} in {{Photocatalyst Design}}},
  author = {Kamat, Prashant V.},
  year = 2012,
  month = mar,
  journal = {J. Phys. Chem. Lett.},
  volume = {3},
  number = {5},
  pages = {663--672},
  publisher = {American Chemical Society},
  doi = {10.1021/jz201629p},
  urldate = {2025-07-16}
}

@article{klumpers2024,
  title = {Transferable, {{Living Data Sets}} for {{Predicting Global Minimum Energy Nanocluster Geometries}}},
  author = {Klumpers, Bart and Hensen, Emiel J. M. and Filot, Ivo A. W.},
  year = 2024,
  month = aug,
  journal = {J. Chem. Theory Comput.},
  volume = {20},
  number = {15},
  pages = {6801--6812},
  publisher = {American Chemical Society},
  issn = {1549-9618},
  doi = {10.1021/acs.jctc.4c00572},
  urldate = {2025-09-17}
}

@article{kresse1993,
  title = {Ab Initio Molecular Dynamics for Liquid Metals},
  author = {Kresse, G. and Hafner, J.},
  year = 1993,
  month = jan,
  journal = {Physical Review B},
  volume = {47},
  number = {1},
  pages = {558--561},
  issn = {0163-1829},
  doi = {10.1103/PhysRevB.47.558},
  urldate = {2017-12-14},
  arxiv = {10.1016/0927-0256(96)00008},
  isbn = {0163-1829 (Print)\textbackslash n0163-1829 (Linking)},
  pmid = {9984901}
}

@article{kresse1994,
  title = {Ab Initio Molecular-Dynamics Simulation of the Liquid-Metal--Amorphous-Semiconductor Transition in Germanium},
  author = {Kresse, G. and Hafner, J.},
  year = 1994,
  month = may,
  journal = {Physical Review B},
  volume = {49},
  number = {20},
  pages = {14251--14269},
  issn = {0163-1829},
  doi = {10.1103/PhysRevB.49.14251},
  urldate = {2017-12-14},
  arxiv = {10.1016/0927-0256(96)00008},
  isbn = {0163-1829 (Print)\textbackslash n0163-1829 (Linking)},
  pmid = {10010505}
}

@article{kresse1996,
  title = {Efficiency of Ab-Initio Total Energy Calculations for Metals and Semiconductors Using a Plane-Wave Basis Set},
  author = {Kresse, G. and Furthm{\"u}ller, J.},
  year = 1996,
  month = jul,
  journal = {Computational Materials Science},
  volume = {6},
  number = {1},
  pages = {15--50},
  publisher = {Elsevier},
  issn = {09270256},
  doi = {10.1016/0927-0256(96)00008-0},
  urldate = {2017-12-14},
  arxiv = {10.1016/0927-0256(96)00008},
  isbn = {0927-0256},
  pmid = {9984901}
}

@article{kresse1996a,
  title = {Efficient Iterative Schemes for Ab Initio Total-Energy Calculations Using a Plane-Wave Basis Set},
  author = {Kresse, G. and Furthm{\"u}ller, J.},
  year = 1996,
  month = oct,
  journal = {Physical Review B},
  volume = {54},
  number = {16},
  pages = {11169--11186},
  publisher = {American Physical Society},
  issn = {0163-1829},
  doi = {10.1103/PhysRevB.54.11169},
  urldate = {2017-12-14},
  arxiv = {10.1016/0927-0256(96)00008},
  isbn = {1098-0121},
  pmid = {9984901}
}

@article{Kresse1999a,
  title = {From Ultrasoft Pseudopotentials to the Projector Augmented-Wave Method},
  author = {Kresse, G. and Joubert, D.},
  year = 1999,
  month = jan,
  journal = {Physical Review B},
  volume = {59},
  number = {3},
  pages = {1758--1775},
  publisher = {American Physical Society},
  issn = {0163-1829},
  doi = {10.1103/PhysRevB.59.1758},
  urldate = {2017-12-14},
  arxiv = {10.1016/0927-0256(96)00008},
  isbn = {0163-1829},
  pmid = {19309091}
}

@article{krishnan2017,
  title = {Investigation of {{Ligand-Stabilized Gold Clusters}} on {{Defect-Rich Titania}}},
  author = {Krishnan, Gowri and Al Qahtani, Hassan S. and Li, Junda and Yin, Yanting and Eom, Namsoon and Golovko, Vladimir B. and Metha, Gregory F. and Andersson, Gunther G.},
  year = 2017,
  month = dec,
  journal = {J. Phys. Chem. C},
  volume = {121},
  number = {50},
  pages = {28007--28016},
  issn = {1932-7447, 1932-7455},
  doi = {10.1021/acs.jpcc.7b09514},
  urldate = {2024-04-12},
  langid = {english}
}

@article{li2017,
  title = {Photocatalytic {{Water Splitting}} on {{Semiconductor-Based Photocatalysts}}},
  author = {Li, Rengui and Li, Can},
  year = 2017,
  journal = {Advances in Catalysis},
  volume = {60},
  pages = {1--57},
  publisher = {Elsevier},
  doi = {10.1016/bs.acat.2017.09.001},
  urldate = {2025-12-17},
  copyright = {https://www.elsevier.com/tdm/userlicense/1.0/},
  langid = {english}
}

@article{li2019,
  title = {Interfacial Effects in Supported Catalysts for Electrocatalysis},
  author = {Li, Hao and Chen, Chen and Yan, Dafeng and Wang, Yanyong and Chen, Ru and Zou, Yuqin and Wang, Shuangyin},
  year = 2019,
  journal = {J. Mater. Chem. A},
  volume = {7},
  number = {41},
  pages = {23432--23450},
  issn = {2050-7488, 2050-7496},
  doi = {10.1039/C9TA04888J},
  urldate = {2025-09-05},
  langid = {english}
}

@article{lin2024,
  title = {Density-{{Functional Theory Studies}} on {{Photocatalysis}} and {{Photoelectrocatalysis}}: {{Challenges}} and {{Opportunities}}},
  shorttitle = {Density-{{Functional Theory Studies}} on {{Photocatalysis}} and {{Photoelectrocatalysis}}},
  author = {Lin, Chun-Han and Rohilla, Jyoti and Kuo, Hsuan-Hung and Chen, Chun-Yi and Mark Chang, Tso-Fu and Sone, Masato and Ingole, Pravin P. and Lo, Yu-Chieh and Hsu, Yung-Jung},
  year = 2024,
  journal = {Solar RRL},
  volume = {8},
  number = {10},
  pages = {2300948},
  issn = {2367-198X},
  doi = {10.1002/solr.202300948},
  urldate = {2025-12-18},
  copyright = {\copyright{} 2024 The Author(s). Solar RRL published by Wiley-VCH GmbH},
  langid = {english}
}

@article{liu2024,
  title = {Atomically {{Precise Metal Nanoclusters}} for {{Photocatalytic Water Splitting}}},
  author = {Liu, Ye and Wang, Yu and Pinna, Nicola},
  year = 2024,
  month = jul,
  journal = {ACS Materials Lett.},
  volume = {6},
  number = {7},
  pages = {2995--3006},
  issn = {2639-4979, 2639-4979},
  doi = {10.1021/acsmaterialslett.4c00622},
  urldate = {2026-01-03},
  copyright = {https://creativecommons.org/licenses/by/4.0/},
  langid = {english}
}

@article{mai2022,
  title = {Machine {{Learning}} for {{Electrocatalyst}} and {{Photocatalyst Design}} and {{Discovery}}},
  author = {Mai, Haoxin and Le, Tu C. and Chen, Dehong and Winkler, David A. and Caruso, Rachel A.},
  year = 2022,
  month = aug,
  journal = {Chem. Rev.},
  volume = {122},
  number = {16},
  pages = {13478--13515},
  publisher = {American Chemical Society},
  issn = {0009-2665},
  doi = {10.1021/acs.chemrev.2c00061},
  urldate = {2025-12-18}
}

@article{manna2023,
  title = {A Database of Low-Energy Atomically Precise Nanoclusters},
  author = {Manna, Sukriti and Wang, Yunzhe and Hernandez, Alberto and Lile, Peter and Liu, Shanping and Mueller, Tim},
  year = 2023,
  month = may,
  journal = {Sci Data},
  volume = {10},
  number = {1},
  pages = {308},
  publisher = {Nature Publishing Group},
  issn = {2052-4463},
  doi = {10.1038/s41597-023-02200-4},
  urldate = {2025-09-22},
  copyright = {2023 The Author(s)},
  langid = {english}
}

@article{medford2015,
  title = {From the {{Sabatier}} Principle to a Predictive Theory of Transition-Metal Heterogeneous Catalysis},
  author = {Medford, Andrew J. and Vojvodic, Aleksandra and Hummelsh{\o}j, Jens S. and Voss, Johannes and {Abild-Pedersen}, Frank and Studt, Felix and Bligaard, Thomas and Nilsson, Anders and N{\o}rskov, Jens K.},
  year = 2015,
  month = aug,
  journal = {Journal of Catalysis},
  series = {Special {{Issue}}: {{The Impact}} of {{Haldor Tops\o e}} on {{Catalysis}}},
  volume = {328},
  pages = {36--42},
  issn = {0021-9517},
  doi = {10.1016/j.jcat.2014.12.033},
  urldate = {2025-12-20}
}

@article{mou2023,
  title = {Bridging the Complexity Gap in Computational Heterogeneous Catalysis with Machine Learning},
  author = {Mou, Tianyou and Pillai, Hemanth Somarajan and Wang, Siwen and Wan, Mingyu and Han, Xue and Schweitzer, Neil M. and Che, Fanglin and Xin, Hongliang},
  year = 2023,
  month = feb,
  journal = {Nat Catal},
  volume = {6},
  number = {2},
  pages = {122--136},
  publisher = {Nature Publishing Group},
  issn = {2520-1158},
  doi = {10.1038/s41929-023-00911-w},
  urldate = {2025-09-29},
  copyright = {2023 Springer Nature Limited},
  langid = {english}
}

@article{norskov2005,
  title = {Trends in the {{Exchange Current}} for {{Hydrogen Evolution}}},
  author = {N{\o}rskov, J. K. and Bligaard, T. and Logadottir, A. and Kitchin, J. R. and Chen, J. G. and Pandelov, S. and Stimming, U.},
  year = 2005,
  month = jan,
  journal = {J. Electrochem. Soc.},
  volume = {152},
  number = {3},
  pages = {J23},
  publisher = {IOP Publishing},
  issn = {1945-7111},
  doi = {10.1149/1.1856988},
  urldate = {2025-12-20},
  langid = {english}
}

@article{norskov2011,
  title = {Density Functional Theory in Surface Chemistry and Catalysis},
  author = {N{\o}rskov, Jens K. and {Abild-Pedersen}, Frank and Studt, Felix and Bligaard, Thomas},
  year = 2011,
  month = jan,
  journal = {Proceedings of the National Academy of Sciences},
  volume = {108},
  number = {3},
  pages = {937--943},
  publisher = {Proceedings of the National Academy of Sciences},
  doi = {10.1073/pnas.1006652108},
  urldate = {2025-12-20}
}

@article{Ong2013,
  title = {Python {{Materials Genomics}} (Pymatgen): {{A}} Robust, Open-Source Python Library for Materials Analysis},
  author = {Ong, Shyue Ping and Richards, William Davidson and Jain, Anubhav and Hautier, Geoffroy and Kocher, Michael and Cholia, Shreyas and Gunter, Dan and Chevrier, Vincent L. and Persson, Kristin A. and Ceder, Gerbrand},
  year = 2013,
  month = feb,
  journal = {Computational Materials Science},
  volume = {68},
  pages = {314--319},
  issn = {09270256},
  doi = {10.1016/j.commatsci.2012.10.028}
}

@article{Perdew1996,
  title = {Generalized {{Gradient Approximation Made Simple}}},
  author = {Perdew, John P. and Burke, Kieron and Ernzerhof, Matthias},
  year = 1996,
  month = oct,
  journal = {Physical Review Letters},
  volume = {77},
  number = {18},
  pages = {3865--3868},
  publisher = {American Physical Society},
  issn = {0031-9007},
  doi = {10.1103/PhysRevLett.77.3865},
  urldate = {2017-12-14},
  arxiv = {10.1016/0927-0256(96)00008},
  isbn = {9780596529321},
  pmid = {10062328}
}

@article{pinaud2013,
  title = {Technical and Economic Feasibility of Centralized Facilities for Solar Hydrogen Production via Photocatalysis and Photoelectrochemistry},
  author = {Pinaud, Blaise A. and Benck, Jesse D. and Seitz, Linsey C. and Forman, Arnold J. and Chen, Zhebo and Deutsch, Todd G. and James, Brian D. and Baum, Kevin N. and Baum, George N. and Ardo, Shane and Wang, Heli and Miller, Eric and Jaramillo, Thomas F.},
  year = 2013,
  journal = {Energy Environ. Sci.},
  volume = {6},
  number = {7},
  pages = {1983},
  issn = {1754-5692, 1754-5706},
  doi = {10.1039/c3ee40831k},
  urldate = {2025-12-17},
  langid = {english}
}

@article{raymondt.tung2014,
  title = {The Physics and Chemistry of the {{Schottky}} Barrier Height},
  author = {{Raymond T. Tung}},
  year = 2014,
  month = jan,
  journal = {Appl. Phys. Rev.},
  volume = {1},
  number = {1},
  pages = {011304},
  issn = {1931-9401},
  doi = {10.1063/1.4858400},
  urldate = {2025-10-21}
}

@article{rodriguez2014,
  title = {Design and Cost Considerations for Practical Solar-Hydrogen Generators},
  author = {Rodriguez, Claudia A. and Modestino, Miguel A. and Psaltis, Demetri and Moser, Christophe},
  year = 2014,
  journal = {Energy Environ. Sci.},
  volume = {7},
  number = {12},
  pages = {3828--3835},
  issn = {1754-5692, 1754-5706},
  doi = {10.1039/C4EE01453G},
  urldate = {2025-12-17},
  langid = {english}
}

@article{samanta2022,
  title = {Challenges of Modeling Nanostructured Materials for Photocatalytic Water Splitting},
  author = {Samanta, Bipasa and {Morales-Garc{\'i}a}, {\'A}ngel and Illas, Francesc and Goga, Nicolae and Antonio~Anta, Juan and Calero, Sofia and {Bieberle-H{\"u}tter}, Anja and Libisch, Florian and {B.~Mu{\~n}oz-Garc{\'i}a}, Ana and Pavone, Michele and Toroker, Maytal Caspary},
  year = 2022,
  journal = {Chemical Society Reviews},
  volume = {51},
  number = {9},
  pages = {3794--3818},
  publisher = {Royal Society of Chemistry},
  doi = {10.1039/D1CS00648G},
  urldate = {2025-12-09},
  langid = {english}
}

@article{shi2021,
  title = {Dynamics of {{Heterogeneous Catalytic Processes}} at {{Operando Conditions}}},
  author = {Shi, Xiangcheng and Lin, Xiaoyun and Luo, Ran and Wu, Shican and Li, Lulu and Zhao, Zhi-Jian and Gong, Jinlong},
  year = 2021,
  month = dec,
  journal = {JACS Au},
  volume = {1},
  number = {12},
  pages = {2100--2120},
  issn = {2691-3704, 2691-3704},
  doi = {10.1021/jacsau.1c00355},
  urldate = {2025-10-09},
  copyright = {https://creativecommons.org/licenses/by-nc-nd/4.0/},
  langid = {english}
}

@article{shirani2023,
  title = {Machine {{Learning Based Electronic Structure Predictors}} in {{Single-Atom Alloys}}: {{A Model Study}} of {{CO Kink-Site Adsorption}} across {{Transition Metal Substrates}}},
  author = {Shirani, Javad and Pham, Hanh D. M. and Yuan, Shuaishuai and Tchagang, Alain B. and Vald{\'e}s, Julio J. and Bevan, Kirk H.},
  year = 2023,
  month = jun,
  journal = {The Journal of Physical Chemistry C},
  volume = {127},
  number = {25},
  pages = {12055--12067},
  issn = {1932-7447},
  doi = {10.1021/acs.jpcc.3c02705}
}

@article{skachkov2021,
  title = {First-Principles Theory for {{Schottky}} Barrier Physics},
  author = {Skachkov, Dmitry and Liu, Shuang-Long and Wang, Yan and Zhang, Xiao-Guang and Cheng, Hai-Ping},
  year = 2021,
  month = jul,
  journal = {Phys. Rev. B},
  volume = {104},
  number = {4},
  pages = {045429},
  issn = {2469-9950, 2469-9969},
  doi = {10.1103/PhysRevB.104.045429},
  urldate = {2025-10-20},
  langid = {english}
}

@article{toyao2020,
  title = {Machine {{Learning}} for {{Catalysis Informatics}}: {{Recent Applications}} and {{Prospects}}},
  shorttitle = {Machine {{Learning}} for {{Catalysis Informatics}}},
  author = {Toyao, Takashi and Maeno, Zen and Takakusagi, Satoru and Kamachi, Takashi and Takigawa, Ichigaku and Shimizu, Ken-ichi},
  year = 2020,
  month = feb,
  journal = {ACS Catal.},
  volume = {10},
  number = {3},
  pages = {2260--2297},
  issn = {2155-5435, 2155-5435},
  doi = {10.1021/acscatal.9b04186},
  urldate = {2025-12-18},
  copyright = {https://doi.org/10.15223/policy-029},
  langid = {english}
}

@article{vandeelen2019,
  title = {Control of Metal-Support Interactions in Heterogeneous Catalysts to Enhance Activity and Selectivity},
  author = {Van Deelen, Tom W. and Hern{\'a}ndez Mej{\'i}a, Carlos and De Jong, Krijn P.},
  year = 2019,
  month = nov,
  journal = {Nat Catal},
  volume = {2},
  number = {11},
  pages = {955--970},
  issn = {2520-1158},
  doi = {10.1038/s41929-019-0364-x},
  urldate = {2025-09-08},
  langid = {english}
}

@article{vandewalle2013,
  title = {Efficient Stochastic Generation of Special Quasirandom Structures},
  author = {Van De Walle, A. and Tiwary, P. and De Jong, M. and Olmsted, D.L. and Asta, M. and Dick, A. and Shin, D. and Wang, Y. and Chen, L.-Q. and Liu, Z.-K.},
  year = 2013,
  month = sep,
  journal = {Calphad},
  volume = {42},
  pages = {13--18},
  issn = {03645916},
  doi = {10.1016/j.calphad.2013.06.006},
  urldate = {2025-01-17},
  langid = {english}
}

@article{wang2019,
  title = {A Quadruple-Band Metal--Nitride Nanowire Artificial Photosynthesis System for High Efficiency Photocatalytic Overall Solar Water Splitting},
  author = {Wang, Yongjie and Wu, Yuanpeng and Sun, Kai and Mi, Zetian},
  year = 2019,
  journal = {Mater. Horiz.},
  volume = {6},
  number = {7},
  pages = {1454--1462},
  issn = {2051-6347, 2051-6355},
  doi = {10.1039/C9MH00257J},
  urldate = {2024-06-03},
  langid = {english}
}

@article{wei1990,
  title = {Electronic Properties of Random Alloys: {{Special}} Quasirandom Structures},
  shorttitle = {Electronic Properties of Random Alloys},
  author = {Wei, S.-H. and Ferreira, L. G. and Bernard, James E. and Zunger, Alex},
  year = 1990,
  month = nov,
  journal = {Phys. Rev. B},
  volume = {42},
  number = {15},
  pages = {9622--9649},
  issn = {0163-1829, 1095-3795},
  doi = {10.1103/PhysRevB.42.9622},
  urldate = {2025-01-17},
  copyright = {http://link.aps.org/licenses/aps-default-license},
  langid = {english}
}

@article{xiao2022,
  title = {Crystallographic {{Effects}} of {{GaN Nanostructures}} in {{Photoelectrochemical Reaction}}},
  author = {Xiao, Yixin and Vanka, Srinivas and Pham, Tuan Anh and Dong, Wan Jae and Sun, Yi and Liu, Xianhe and Navid, Ishtiaque Ahmed and Varley, Joel B. and Hajibabaei, Hamed and Hamann, Thomas W. and Ogitsu, Tadashi and Mi, Zetian},
  year = 2022,
  month = mar,
  journal = {Nano Lett.},
  volume = {22},
  number = {6},
  pages = {2236--2243},
  issn = {1530-6984, 1530-6992},
  doi = {10.1021/acs.nanolett.1c04220},
  urldate = {2024-06-03},
  copyright = {https://doi.org/10.15223/policy-029},
  langid = {english}
}

@article{xiao2023,
  title = {Oxynitrides Enabled Photoelectrochemical Water Splitting with over 3,000 Hrs Stable Operation in Practical Two-Electrode Configuration},
  author = {Xiao, Yixin and Kong, Xianghua and Vanka, Srinivas and Dong, Wan Jae and Zeng, Guosong and Ye, Zhengwei and Sun, Kai and Navid, Ishtiaque Ahmed and Zhou, Baowen and Toma, Francesca M. and Guo, Hong and Mi, Zetian},
  year = 2023,
  month = apr,
  journal = {Nat Commun},
  volume = {14},
  number = {1},
  pages = {2047},
  issn = {2041-1723},
  doi = {10.1038/s41467-023-37754-9},
  urldate = {2025-06-18},
  langid = {english}
}

@article{xu2023,
  title = {Design of the {{Synergistic Rectifying Interfaces}} in {{Mott}}--{{Schottky Catalysts}}},
  author = {Xu, Dong and Zhang, Shi-Nan and Chen, Jie-Sheng and Li, Xin-Hao},
  year = 2023,
  month = jan,
  journal = {Chem. Rev.},
  volume = {123},
  number = {1},
  pages = {1--30},
  publisher = {American Chemical Society},
  issn = {0009-2665},
  doi = {10.1021/acs.chemrev.2c00426},
  urldate = {2025-07-16}
}

@article{xu2024,
  title = {Revisiting the Universal Principle for the Rational Design of Single-Atom Electrocatalysts},
  author = {Xu, Haoxiang and Cheng, Daojian and Cao, Dapeng and Zeng, Xiao Cheng},
  year = 2024,
  month = feb,
  journal = {Nat Catal},
  volume = {7},
  number = {2},
  pages = {207--218},
  issn = {2520-1158},
  doi = {10.1038/s41929-023-01106-z},
  urldate = {2025-02-10},
  langid = {english}
}

@article{zhang2012,
  title = {Band {{Bending}} in {{Semiconductors}}: {{Chemical}} and {{Physical Consequences}} at {{Surfaces}} and {{Interfaces}}},
  shorttitle = {Band {{Bending}} in {{Semiconductors}}},
  author = {Zhang, Zhen and Yates, John T.},
  year = 2012,
  month = oct,
  journal = {Chem. Rev.},
  volume = {112},
  number = {10},
  pages = {5520--5551},
  issn = {0009-2665, 1520-6890},
  doi = {10.1021/cr3000626},
  urldate = {2024-11-07},
  langid = {english}
}

@article{zhong2020,
  title = {Surface Morphology of Polar, Semipolar and Nonpolar Freestanding {{GaN}} after Chemical Etching},
  author = {Zhong, Haijian and Zhang, Chunyu and Song, Wentao and Chen, Kebei and Sheng, Yaohuan and Xu, Gengzhao and Liu, Zhenghui},
  year = 2020,
  month = may,
  journal = {Applied Surface Science},
  volume = {511},
  pages = {145524},
  issn = {01694332},
  doi = {10.1016/j.apsusc.2020.145524},
  urldate = {2025-06-11},
  langid = {english}
}

@article{zhou2023,
  title = {Solar-to-Hydrogen Efficiency of More than 9\% in Photocatalytic Water Splitting},
  author = {Zhou, Peng and Navid, Ishtiaque Ahmed and Ma, Yongjin and Xiao, Yixin and Wang, Ping and Ye, Zhengwei and Zhou, Baowen and Sun, Kai and Mi, Zetian},
  year = 2023,
  month = jan,
  journal = {Nature},
  volume = {613},
  number = {7942},
  pages = {66--70},
  issn = {0028-0836, 1476-4687},
  doi = {10.1038/s41586-022-05399-1},
  urldate = {2024-04-18},
  langid = {english}
}

@incollection{zhou2024,
  title = {Gallium {{Nitride}}-{{Based Artificial Photosynthesis Integrated Devices}} for {{Solar Hydrogen Generation}} and {{Carbon Dioxide Reduction}}},
  booktitle = {Conversion of {{Water}} and {{CO2}} to {{Fuels}} Using {{Solar Energy}}},
  author = {Zhou, Baowen and Zhou, Peng and Dong, Wanjae and Mi, Zetian},
  editor = {Varghese, Oomman K. and Souza, Flavio L.},
  year = 2024,
  month = feb,
  edition = {1},
  pages = {309--339},
  publisher = {Wiley},
  doi = {10.1002/9781119600862.ch11},
  urldate = {2024-05-23},
  isbn = {978-1-119-60084-8 978-1-119-60086-2},
  langid = {english}
}

@article{zunger1990,
  title = {Special Quasirandom Structures},
  author = {Zunger, Alex and Wei, S.-H. and Ferreira, L. G. and Bernard, James E.},
  year = 1990,
  month = jul,
  journal = {Phys. Rev. Lett.},
  volume = {65},
  number = {3},
  pages = {353--356},
  issn = {0031-9007},
  doi = {10.1103/PhysRevLett.65.353},
  urldate = {2025-01-17},
  copyright = {http://link.aps.org/licenses/aps-default-license},
  langid = {english}
}
\end{document}